\documentclass[%
preprint,
https://www.overleaf.com/project/600e228b7b9acb7bce612df0
amsmath,amssymb,
]{revtex4-1}
\usepackage{comment}
\usepackage{tabularx}
\usepackage[normalem]{ulem} 
\usepackage{xcolor, soul} 
\sethlcolor{green}
\usepackage{soul}
\usepackage{graphicx}
\usepackage{mathrsfs}
\usepackage{amsmath}
\usepackage{placeins}
\usepackage{epstopdf, epsfig}
\usepackage{subcaption}
\usepackage[export]{adjustbox}
\usepackage[%
colorlinks=true,
pdfborder={0 0 0},
linkcolor=blue,
citecolor=blue,
urlcolor=blue
]{hyperref}

\usepackage{dcolumn}
\usepackage{bm}
\newcommand{\upi}{\pi}

\begin{document}





\title{Dimension Reduced Turbulent Flow Data From Deep Vector Quantizers}

	\author{Mohammadreza Momenifar}
	\email{mohammadreza.momenifar@duke.edu}
	\author{Enmao Diao}
	\email{enmao.diao@duke.edu}
	\author{Vahid Tarokh}
	\email{vahid.tarokh@duke.edu}
	\author{Andrew D. Bragg}
	\email{andrew.bragg@duke.edu}
	\affiliation{Department of Civil and Environmental Engineering}
	\affiliation{Department of Electrical and Computer Engineering, Duke University, Durham, North Carolina 27708, USA}

\begin{abstract}
	Analyzing large-scale data from simulations of turbulent flows is memory intensive, requiring significant resources. This major challenge highlights the need for data compression techniques. In this study, we apply a  physics-informed Deep Learning technique based on vector quantization to generate a discrete, low-dimensional representation of data from simulations of three-dimensional turbulent flows. The deep learning framework is composed of convolutional layers and incorporates physical constraints on the flow, such as preserving incompressibility and global statistical characteristics of the velocity gradients. The accuracy of the model is assessed using statistical, comparison-based similarity and physics-based metrics. The training data set is produced from Direct Numerical Simulation of an incompressible, statistically stationary, isotropic turbulent flow. 
	The performance of this lossy data compression scheme is evaluated not only with unseen data from the stationary, isotropic turbulent flow, but also with data from decaying isotropic turbulence, a Taylor-Green vortex flow, and a turbulent channel flow. Defining the compression ratio (CR) as the ratio of original data size to the compressed one, the results show that our model based on vector quantization can offer CR$=85$ with a mean square error (MSE) of $O(10^{-3})$, and predictions that faithfully reproduce the statistics of the flow, except at the very smallest scales where there is some loss. Compared to the recent study of Glaws. et. al. (Physical Review Fluids, 5(11):114602, 2020), which was based on a conventional autoencoder (where compression is performed in a continuous space), our model improves the CR by more than $30$ percent, and reduces the MSE by an order of magnitude. Our compression model is an attractive solution for situations where fast, high quality and low-overhead encoding and decoding of large data are required. 

\end{abstract}

\maketitle

\vskip -0.5 cm
\section{Introduction}
\vskip -0.5 cm
Turbulence is a complex dynamical system which is high-dimensional, multi-scale, non-linear, and non-local, exhibiting spatio-temporal chaotic interactions among its very wide range of scales. This has attracted substantial interest for many years both because of the intellectually stimulating challenges associated with its understanding, and also because of its practical importance to a wide range of applications. Examples of such applications include those involving the motion of particles in turbulent flows (\cite{momenifar2018influence,momenifar2020local}) which are critical for environmental sciences (such as atmospheric pollution transport \cite{maxey1987gravitational}, cloud formation \cite{beard1993warm}, volcanic eruptions \cite{ongaro2007parallel}, orographic rainfall \cite{eghdami2019extreme}) and combustion processes \cite{kuo2012fundamentals}, and the effect of turbulence on the design and performance of engineering devices (such as wind turbines \cite{abdulqadir2017physical, moriarty2002effect}, airplane design \cite{etkin1981turbulent}, heat exchangers \cite{bianco2011numerical,momenifar2015effect,nasr2015heat}, and pumps \cite{fan2011computational, hanafizadeh2014void}), to name just a few.  

The growing availability of computational resources in recent decades has facilitated the exploration and understanding of turbulent flows. From the Computational Fluid Dynamics (CFD) perspective, high-fidelity simulations of turbulent flows may be achieved by solving the Navier-Stokes (NS) equation numerically, a procedure known as Direct Numerical Simulation (DNS). A less expensive computational  approach is to use Large Eddy simulation (LES) in which the small scales are modeled rather than simulated, reducing computational cost at the expense of accuracy and access to small-scale flow information. Both DNS and LES require significant computational resources, and can produce large amounts of data that is cumbersome to store, transfer (bandwidth requirements) and analyze. Such challenges can be addressed by the development of efficient and accurate data compression techniques.   
\par \smallskip
Data compression in this CFD context may be employed for both compressed data checkpointing (storing the compressed data and using it for restarting the simulation), and post-processing purposes.   
The main goal of such compression processes is to truncate the size of the data files while also ensuring that (i) simulation restarts using the compressed data would not significantly impact the long-term behavior of the simulated flow (ii) the compressed data still preserves the essential statistical properties of the turbulent flow with reasonable accuracy. 
Compression techniques can be broadly classified into lossless and lossy, depending on the information loss during a data reduction process. In dealing with large CFD simulations, often a lossy compression algorithm is favored, due to its reduced memory requirements while offering higher compression capabilities, although this approach sacrifices the ability to retrieve the original data. Recent studies in the image compression community (\cite{diao2020drasic,johnston2018improved}) have shown that machine-learning based lossy compression techniques, particularly those based on deep neural networks, can outperform classical methods such as JPEG standard \cite{wallace1992jpeg} and BPG \cite{bellard2015bpg}. 
\par \smallskip
Machine learning techniques are bringing fresh perspectives in many areas. Among machine learning techniques, deep learning models have received significant attention due to their ability to capture complex interactions and achieve outstanding performance across a wide range of applications in information technology, healthcare and engineering, to name a few. In the context of data compression, common deep neural networks are
based on an autoencoder architecture, which has also gained interest in the CFD community for the purpose of data compression of three dimensional turbulent flows (\cite{glaws2020deep}, \cite{mohan2020spatio}).
Our study is in line with these recent endeavors and proposes a deep learning based data compression framework that compresses/encodes the velocity field data of three dimensional turbulent flows into a discrete space through quantization of the bottleneck representation, the so-called vector-quantized autoencoder. This approach involves a calibration process that enables infusing prior knowledge of the data to boost the performance of model. Compared to the recent autoencoder network proposed in \cite{glaws2020deep}, training our framework is computationally much cheaper and the trained model significantly improves not only the compression ratio but also the accuracy of the reconstructed data. 
\par \smallskip
This outline of this paper is given next. A  brief literature review on the data compression techniques, application of deep learning in turbulence research and data compression is provided in Section \ref{Literature}. In Sections \ref{Methodology} our mathematical methodology, and in  Section \ref{Computational_Details} our implementation approach is discussed.  Assessment metrics and methodologies are described in Section \ref{Evaluation_Metrics}. Using these metrics, numerical results are provided  in Section \ref{Results_Discussion} evaluating the performance of our model on several test cases. These test cases represent turbulent flows with different characteristics from the training set data. Finally, Section \ref{Conclusions} provides a brief summary of our findings.
\vskip -0.5 cm
\section{Background}\label{Literature}
\vskip -0.5 cm
\subsection{Data Compression}
As scale of high-fidelity turbulent flow simulations grows and becomes data intensive, data compression becomes an integral part to reduce burdens on computational resources, in terms of disk usage and memory bandwidth, during data management which includes storage, transfer, and in-situ analysis of data. While lossless compression technologies can eliminate less useful content of data without affecting the quality, their compression capabilities are modest/limited \cite{engelson2000lossless,ratanaworabhan2006fast} and cannot drastically reduce the size of data files. On the other hand, lossy schemes can significantly reduce the size of data files while incurring some distortion in the reconstructed data when decoding the compressed file. Seven use cases of lossy compression for scientific research are discussed in \cite{cappello2019use} which are visualization, reducing data stream intensity, reducing footprint on storage, reducing I/O time, accelerating checkpoint/restart, reducing footprint in memory, and accelerating execution. 
The analysis of large turbulent flow simulations is generally more concentrated on the statistical quantities of flow field rather than its instantaneous values. Therefore, we prefer a lossy compression scheme that can extract the most relevant physics of turbulence hence resulting in a minimal impact on the quality of statistics of post-processed data from a physics point of view. Given that loss of information is unavoidable in lossy compression methods, we desire a model that can offer a controllable level of trade-off between the compression ratio and distortion of data, independent of input data. 

Lossy compression schemes involve two steps of (i) compression, in which original data is encoded into a latent space and (ii) decompression, in which the compressed data is decoded. The quality of these lossy dimension reduction models is evaluated based on their compression capability, which is measured by computing compression ratio (CR) and defined as :
\begin{eqnarray}\label{eq:CR}
CR = \frac{original \: file \: size}{compressed \: file \: size}
\end{eqnarray}
and their accuracy in reconstructing the original data, which is measured by different metrics depending on the applications. 
Lossy compression schemes, which are directly inherited from image compression techniques (\cite{cappello2019use}), cover a wide range of approaches such as
classical and diffusion wavelets (\cite{coifman2006diffusion,mallat1989theory}), spectral methods (\cite{karni2000spectral}), reduced-order models (\cite{blanc2014analysis}), orthogonal transforms (\cite{rao2014discrete}) and matrix decomposition (\cite{cheng2005compression}). In the computer vision community, many recent studies, such as \cite{toderici2015variable, mentzer2018conditional}, reported that application of deep neural networks in the problem of lossy image compression can yield comparable results as classical techniques (JPEG \cite{wallace1992jpeg} and its variants such as BPG \cite{bellard2015bpg}) and even \cite{johnston2018improved} reported that their deep learning based lossy image compression framework can outperform them.
In the following sections, we discuss deep neural networks and their applications in data compression, particularly for turbulence research.  


\subsection{Deep Learning in Turbulence Research}
With the growing availability of high-fidelity data and computational power, data-driven modeling has recently gained enormous interest. Among these data-driven methods, Machine Learning (ML) techniques, particularly deep learning models, have received much attention due to their ability to capture complex interactions, and achieve outstanding performance across a wide range of applications in information technology, healthcare and engineering, to name a few.
Deep learning models, also known as deep neural networks, are built from a stack of layers with learnable parameters that perform linear and non-linear transformation between their inputs and outputs where their parameters are trained through an optimization procedure (\cite{lecun2015deep}), referred to as backpropagation.

Mathematically speaking, Artificial Neural Networks (ANN), including deep learning, can be viewed as generalization of linear models (\cite{muller2016introduction}). In an ANN, the basic unit is a neuron (also called node) and represents a vector-to-scaler function. A layer is composed of a set of such nodes and operates as a vector-to-vector function. The notion of being “deep” refers to the interconnected nature of the stacked layers (also known as fully connected neural networks), in which the output of each layer ($\boldsymbol{x}$) passes through an activation function ($\phi$) and becomes the input of the subsequent layer ($\boldsymbol{y}= \phi(\boldsymbol{W}^\top \boldsymbol{x})$, where $\boldsymbol{W}$ are the weights), enabling the construction of a universal function approximator \cite{ hornik1989multilayer, hornik1990universal}. Due to this structure, the model can describe complex nonlinear functions (although there is no guarantee that it would be able to learn such a function), with the model parameters and weights $\boldsymbol{W}$ learned during training process. Interested readers are referred to \cite{goodfellow2016deep} for more detailed information.


Despite the remarkable success of Deep Neural Networks and their ability to outperform other competitive algorithms in many areas of big data analytics, their applications in the physical sciences has been tempered by the caveat that these networks are essentially ``black boxes'' which are hard to interpret and physics-agnostic. Furthermore, training such networks requires significant data (except for unsupervised learning tasks), which may not be available, and the training process takes some time. However, well-trained deep learning models can make quick inferences about similar input data, particularly for very complex physical systems, and this characteristic distinguishes them from conventional numerical solvers.

In the fluid dynamics community, particularly CFD of turbulent flows, there have been attempts in recent years to utilize deep learning tools for turbulence modeling. Such endeavors have mainly focussed on developing Reynolds stress closures and subgrid-scale (SGS) models for RANS and LES simulations, respectively (\cite{maulik2019subgrid,ling2016reynolds,beck2019deep}), super-resolution reconstruction of coarse flow fields (\cite{fukami2018super}), augmenting existing turbulence models with physics-informed machine learning (\cite{wu2018physics,wang2017origin}), and spatio-temporal modeling of isotropic turbulence \cite{mohan2020spatio}.

The first step of the modeling process is to identify potential deep learning models which are well-suited for handling non-linear spatial  data. Indeed, we seek models that can capture the spatial dependencies of turbulent flow fields. Given the structured grid/pixel-based discritizations of many computational domains in turbulent flow simulations, we can draw an analogy between turbulence data and image data where three components of velocity field represent RGB (red, green, blue) color channels, and instead of two dimensional spatial grids we have a volumetric data. Therefore, we can benefit from modern computer vision approaches by applying them in a turbulence modeling framework. The recent success of deep learning schemes in vision tasks (image classification, object detection, face recognition, etc.) is heavily attributed to developments in Convolutional Neural Networks (CNN) (\cite{khan2018guide}), which are inspired by the organization of animal visual cortex (\cite{hubel1968receptive,fukushima1982neocognitron}).

Deep convolutional neural networks have a compositional hierarchy that are designed to learn spatial features from low to high level patterns. In the context of image recognition for instance, the first layers capture low-level features such as edges, and subsequent layers assemble edges to form shapes, and finally deeper layers combine these shapes to construct objects (\cite{salehipour2018deep}). CNNs assume that there is a positional relationship in data, like rows and columns of images. Therefore, the key advantage of CNNs over traditional ANNs is that they can account for correlations between the adjacent input data points. 
From the implementation and mathematical perspectives, the weights (learnable parameters) in such networks are structured as kernels (or filters), with a predetermined size, which sweep across structured input data (an image for instance) while performing a convolution operation (that is actually a cross-correlation). These kernels, known as convolutional kernels, are applied to extract a reduced data space from the input data (a feature extraction approach) while preserving the positional relationships between the data points.
Unlike traditional fully connected neural networks, there is no need to have all the neurons be connected to each other, which this leads to a careful pruning of the connections based on the structure of the data. This then reduces the number of weights required to be trained in the network. These characteristics make CNNs more robust to rotation and translation, and to be agnostic to the spatial dimension of the input grid points. Selection of a convolution kernel involves choosing kernel size, stride length and padding. More information on convolutional layers can be found in \cite{goodfellow2016deep}. 

The capability of CNNs in extracting spatial connections in data (i.e. exploiting the correlations between the adjacent input data points) could make them a suitable architecture for a variety of physics-based applications. Furthermore, one can incorporate different boundary conditions (wall, periodic etc.) into convolutional layers in the form of padding (\cite{patil2019development}). 
Such a hierarchical superposition of complex structures may resemble the cascade nature of turbulence wherein there is on average a transfer of information (energy) among the scales, and hence it is worth exploring the extent to which the techniques utilized in computer vision tasks could be applicable to turbulent flow fields. Recent studies of \cite{wang2020towards} and \cite{li2020fourier}) have utilized CNNs in the context of spatio-temporal modeling of 2D turbulent flows and found promising results. In the following section, we discuss the application of CNNs in building lossy data compression schemes and survey the recent relevant works in the turbulence community.

\subsection{Deep Learning for Data Compression}
With the rise of deep learning and the rapid development of its theory, its applications in the field of image compression have proven remarkably successful. Common deep neural networks for the purpose of image compression have an auto-encoder architecture, including recurrent and non-recurrent auto-encoders, and are mainly composed of CNN layers \cite{diao2020drasic}. The auto-encoder architecture is composed of two networks, namely the encoder and decoder. The encoder performs down-sampling on the input data to generate a compressed non-linear latent space (also called the bottleneck) representation of the input, while the decoder takes the output of the encoder and reconstructs the input data via upsampling. 
Unlike fully connected neural networks based autoencoders that receive a vectorized representation of the input data which leads to the removal of dimensional information, fully convolutional autoencoders take original the representation of the data and can maintain the spatial relationships between data points, and are agnostic to the spatial dimensions of the input data. In CNN based autoencoders, dimension reduction can be performed by appropriate choices of the convolution kernel and pooling layers, while the upsampling is performed with transposed convolution kernels.

Convolutional Neural Network-based autoencoder has also received attention in the turbulence community for use in dimension reduction of velocity field data from three-dimensional turbulent flows. In the recent study of \cite{mohan2020spatio}, a convolutional autoencoder was employed to obtain a low dimensional representation of a three-dimensional velocity field from an isotropic turbulent flow. The main purpose of their study was to model spatio-temporal realizations of isotropic turbulence. However, to handle the data, they had to use a data reduction technique to obtain a compact version of the data before passing it to their sequence learning model. They constructed a framework that offers a compression ratio of $125$. Their compression results indicate that their model can capture large scales of flow accurately and inertial scales with some distortion, but it failed drastically in preserving the small scales of flow, as seen in their compression model results for the turbulence energy spectrum and the probability distributions of the longitudinal velocity gradients. 

The recent turbulence data compression study of \cite{glaws2020deep} centered around the outperformance of their CNN based autoencoder against a variant of singular value decomposition (SVD) for the purpose of in-situ data compression, with a focus on generating lossy restart data. Their autoencoder offers a compression ratio of $64$ and has been trained on decaying isotropic turbulence, and then tested on Taylor-Green vortex and pressure-driven channel flows. Their findings clearly demonstrate the remarkable performance of their model in reconstructing physical characteristics of flows (e.g. near-wall behavior in channel flows) which were not seen by the model during training. Their results for the lossy restart data show that the trained network produces compressed data files that, when used as restart files in simulations, leads to simulation results that preserve the important flow properties when compared with the results generated using the original, full data for the restart files. Nevertheless, although their model does a much better job at capturing the properties of the inertial range scales compared to the model in \cite{mohan2020spatio}, it still it leads to a noticeable loss of information for the small scales of flow.

The results of these recent studies indicate that further efforts in using CNN-based autoencoders for data compression of turbulent flows can be very fruitful. Our study is motivated by these recent works to design a data compression model that increase not only the compression ratio but also the accuracy of the reconstructed data. From a theoretical point of view, the main difference between our framework and these studies is that we incorporate the concept of vector quantization and generate a bottleneck representation in a discrete space (an integer representation). Furthermore, we infuse prior physical knowledge of the data into the model to enhance the ability of model in respecting the small-scale characteristics of the turbulence which were not well-captured in previous works. From the implementation point of view, we design and train our framework such that it results in a  significant decrease in the number of hyper-parameters which eventually reduces the computational cost. Furthermore, additional (off-line) training of a spatio-temporal model over such a discrete latent space is $CR \times $ faster, allowing the model to be trained efficiently on higher resolution data.
In section \ref{Methodology}, we elaborate on the theory underlying the model, and in section \ref{Computational_Details} we discuss the details of the implementation and computations. To the best of our knowledge, this is the first deep learning framework embedded with the quantization concept in the turbulence literature. Our results indicate that our framework outperforms the existing models while not suffering from their shortcomings of capturing small scales.


%

\vskip -0.5 cm
\section{Methodology}\label{Methodology}
\vskip -0.5 cm
\subsection{Vector Quantization}
In the context of signal compression, quantization means mapping a real (floating point) value to an integer which is intrinsically lossy. Following this definition, vector \footnote{not limited to the vector-valued data; adjacent pixels in an image can form a vector} quantization (VQ) can be described as clustering vectors with similar values in which all the vectors that fall within a cluster are quantized to have the same (binary) value corresponding to the index of the cluster. Each cluster (a Vorono\text{\"i} region) has a unique vector representation, called codeword, and a mapping scheme is composed of a collection of such clusters, called a codebook. Given that input vectors are most likely non-uniformly distributed in space, thinking of a vector in $\mathbb{R}^{n}$ as a point, a good codebook takes this into account and allocates less codewords to sparse regions. That eventually leads to low error values for commonly occurring points and high error values for the rare ones \cite{li2018data}. 

Mathematically speaking, in a VQ operation, $m$-dimensional vectors in ${\mathbb{R}^m}$ are mapped into a finite set of codewords/vectors
$\mathcal{Y}=\{e_{i}:i=1,2,..,K\}$ with a fixed size $D$ or $e_{i}\in {\mathbb{R}^D}$, where $K$ represents the size of the codebook. Larger $D$ or $K$ can improve the expressibility of the codebook which comes at the cost of reducing the compression ratio. 

A VQ is composed of encoding and decoding operations. In terms of computations, the encoding is a exhaustive task (designing a suitable codebook) while decoding (retrieving the index of a codeword in a codebook that has the minimum Euclidean distance to the input vector) is quite light. Therefore, a VQ is an attractive solution for applications that involve encode-once-and-decode-many times \cite{li2018data}, such as data compression of CFD simulations. 
\subsection{Vector-Quantized Autoencoder}
A Vector-Quantized (fully convolutional) autoencoder encodes the input data in a discrete latent space and can effectively use the capacity of latent space by conserving important features of data that usually span many dimensions in data space (such as objects in images) and reducing entropy (putting less focus on noise) \cite{van2017neural}. Given that turbulence is inherently non-local (its features span many dimensions in the computational domain), VQ might be a good tool to retain import characteristics of turbulence during compression of the data. 

Compared to a conventional autoencoder, a Vector-Quantized Autoencoder has an additional Vector-Quantizer module. The encoder ($E$) serves as a non-linear function that maps input data ($x$) to a vector $E(x)$. The quantizer modules takes this vector and outputs an index ($k$) corresponding to the closest codeword in the codebook to this vector ($e_{k}$):

\begin{eqnarray}\label{eq:Quantizer}
Quantize(E(x)) = e_{k},\, \text{where} \; k = \underset{j}{\arg\min} \parallel E(x) - e_{j} \parallel_{2}.
\end{eqnarray}

Codeword index $k$ is used for the integer representation of the latent space, and $e_{k}$ serves as the input of decoder ($D$) which operates as another non-linear function to reconstruct the input data. The Vector-Quantizer module brings two additional terms in the loss function, namely codebook loss and commitment loss, to align the encoder output with the vector space of the codebook. The entire VQ-AE loss is defined as:

\begin{eqnarray}\label{eq:VQ_VAE_Loss}
\mathcal{L}(x,D(e))=\underbrace{\parallel x - D(e) \parallel^{2}_{2}}_{reconstruction~loss} + \underbrace{\parallel sg\{E(x)\} - e \parallel^{2}_{2}}_{codebook~loss} + \underbrace{\beta \parallel sg\{e\} - E(x) \parallel^{2}_{2}}_{commitment~loss}. 
\end{eqnarray}
\\
In the above, the reconstruction loss trains both encoder and decoder parameters where 
the gradient of reconstruction error is back-propagated to the decoder first, and then directly passed to the output of the encoder using the straight-through gradient estimator \cite{bengio2013estimating} (because there is no real gradient for the \emph{$\arg\min$} term in Equation (\ref{eq:Quantizer})). The codebook loss trains only the codebook by moving the picked codeword $e_{k}$ towards the output of the encoder. The commitment loss trains only the decoder by encouraging the output of the encoder to be close to the selected codeword, and to avoid jumping frequently between different codewords in the codebook. The $\beta$ coefficient represents the commitment coefficient which controls the reluctance against this fluctuation, and $sg\{\cdot\}$ denotes the stop gradient operator which does not propagate gradients with respect to its arguments. For the codebook design, the codebook loss can be replaced with the exponential moving average scheme to update codebook parameters. More details can be found in \cite{van2017neural,razavi2019generating}.

\subsection{Incorporating prior knowledge of data into framework}\label{Incorporating_prior_knowledge}
So far we have highlighted two connections between the characteristics of turbulence and the proposed framework: (i) the compositional hierarchy of convolutional networks and multiscale nature of turbulence. (ii) the capability of VQ to capture spatially correlated features of the flow that also exist between different scales. These are general characteristics of many multi-scale physical phenomenon. Consequently, the above framework can be employed for other applications such as climatology, geophysics, oceanography, astronomy, and astrophysics where the problem of interest is understanding patterns and correlations. We employ this framework for the case of turbulent flows. It would be beneficial if we could infuse our prior knowledge of input data into the model so that it is enforced to obey those constraints. To this end, we impose such constraints by additional regularization terms in the loss function of the model. 

As noted earlier, preserving small-scale properties of the turbulent flow was a challenge for prior compression models.  It may be of interest to add appropriate constraints in order to capture these more faithfully. Given that our model will be trained on isotropic turbulence, the appropriate constraints for this kind of flow will be our main focus here. Let us consider the Cartesian components of the velocity gradient tensor, $A_{ij}=\partial u_i/\partial x_j$. The incompressibility of the flow implies that $A_{ii} = 0$. Furthermore, ``Betchov relations'' \cite{betchov1956inequality} for an incompressible, statistically homogeneous turbulent flow are given by

\begin{eqnarray}\label{eq:Betchov_1}
\langle S_{ij} S_{ij} \rangle = \langle R_{ij} R_{ij} \rangle = \frac{1}{2}\langle \omega_{i} \omega_{i} \rangle \\
\label{eq:Betchov_2}
\langle S_{ik} S_{kj} S_{ij}\rangle = -\frac{3}{4}\langle S_{ij} \omega_{i} \omega_{j} \rangle
\end{eqnarray}

where $S_{ij} \equiv (1/2)(A_{ij}+A_{ji})$ is the strain-rate, $R_{ij} \equiv (1/2)(A_{ij}-A_{ji})$ is the rotation-rate, and $\omega_{i} = \epsilon_{ijk}R_{jk}$is the vorticity (where $\epsilon_{ijk}$ is the Levi-Civita symbol). We summarize all these constrains as:

\begin{eqnarray}\label{eq:VG_Constraint}
\text{Velocity Gradient Constraint (VGC)} = \underbrace{MSE(A_{ij},\widehat{A_{ij}})}_{i= j} + a \times \underbrace{MSE(A_{ij},\widehat{A_{ij}})}_{i\neq j} \\
\label{eq:Other_constraints}
\begin{split}
\text{Other Constraints (OC)} =  MAE(\langle S_{ij} S_{ij} \rangle,\widehat{\langle S_{ij} S_{ij} \rangle}) + 
MAE(\langle R_{ij} R_{ij} \rangle,\widehat{\langle R_{ij} R_{ij} \rangle}) + \\
MAE(\langle S_{ik} S_{kj} S_{ij} \rangle,\widehat{\langle S_{ik} S_{kj} S_{ij} \rangle})+
MAE(\langle S_{ij} \omega_{i} \omega_{j} \rangle,\widehat{\langle S_{ij} \omega_{i} \omega_{j} \rangle}), 
\end{split}
\end{eqnarray}
\noindent where for each quantity of interest, $A_{ij}$,
$\langle S_{ij} S_{ij} \rangle$, $\langle R_{ij} R_{ij} \rangle$, $\langle S_{ik} S_{kj} S_{ij} \rangle$, and $\langle S_{ij} \omega_{i} \omega_{j} \rangle$
the reconstructed ones are denoted by $\widehat{A_{ij}}$, $\widehat{\langle S_{ij} S_{ij} \rangle}$, $\widehat{\langle R_{ij} R_{ij} \rangle}$, $\widehat{\langle S_{ik} S_{kj} S_{ij} \rangle}$, and $\widehat{\langle S_{ij} \omega_{i} \omega_{j} \rangle}$, respectively. The coefficient $a$ is introduced in equation \ref{eq:VG_Constraint} to account for the differences in the statistics of the longitudinal and transverse components. For example, at the small-scales of isotropic turbulence, the variance of the transverse velocity gradients are twice the size of the longitudinal ones \cite{pope_2000}. We penalize the deviations in the Equation (\ref{eq:Other_constraints}) with mean absolute error ($MAE$), which is less sever than the $MSE$ in the Equation (\ref{eq:VG_Constraint}), in order to put less focus on this term as it is a secondary objective (usually the error at high-order statistics are large and we mainly want to focus on recovering the velocity field and its first-order statistics).
Finally adding these constraints as regularization terms to VQ-AE loss function gives the overall loss function (OL) given below
\begin{eqnarray}\label{eq:Final_loss}
\text{Overall Loss (OL) } = \text{VQ-AE loss} + \alpha \times \text{VGC} + \gamma \times \text{OC},
\end{eqnarray}
which will be use for training.

\vskip -0.5 cm
\section{Computational Details}\label{Computational_Details}
\vskip -0.5 cm
\subsection{Training and Testing Data}
We train our model using high-fidelity DNS data of a three-dimensional, statistically stationary, isotropic turbulent flow with Taylor-Reynolds number $R_\lambda = 90$, solved on a cubic domain with $128$ grid points in each direction (DNS 1 in Table~\ref{tab:parameters}). More details on the DNS can be found in our previous works (\cite{ireland2013highly,momenifar2020local,momenifar2018influence}). In our training data set we have $40$ snapshots (as opposed to $3000$ in the recent study of \cite{glaws2020deep}) equally spaced in time, covering a time span of $2T_{l}$ ($T_{l}$ denotes large eddy turn over time). Each snapshot consists of the three components of the velocity field ($u , v, w$) at all $128^3$ grid points.

We test the performance of the trained framework on a variety of flows to determine how well the model can compress flows different from those used in the training. For comparison purposes, the flows are selected to correspond to those considered in \cite{glaws2020deep}. Specifically, we start with statistically stationary isotropic turbulence, but then consider decaying isotropic turbulence, a Taylor-Green vortex flow, and finally a turbulent channel flow which possesses strong anisotropy and inhomogeneity (neither of which are present in the training data). The test snapshot of stationary isotropic turbulence comes from the same DNS that was used for training the model, but this snapshot is from a time that is $2T_{l}$ beyond the time of the last snapshot used in the training set. We obtained the snapshot for decaying isotropic turbulence from the Github repository of \cite{glaws2020deep}, which represents a flow with $R_\lambda = 89$ simulated on a $128^3$ grid. Our Taylor-Green vortex snapshot comes from a DNS with $Re = 1600$ (where $\nu = 1/1600$) performed on a $192^3$ grid, similar to that used in  \cite{glaws2020deep}. Finally, our channel flow snapshot comes from a DNS using a $2\pi \: \times 2 \: \times \pi$ domain with friction Reynolds number $Re_{\tau} = 180$, performed on a $128 \times 129 \times 128$ grid.

%
\begin{table}
	\centering
	\renewcommand{\arraystretch}{0.85}
	\setlength{\tabcolsep}{12pt}
	\begin{tabularx}{1\textwidth}{| >{\centering\arraybackslash}X
  | >{\centering\arraybackslash}X 
  | >{\centering\arraybackslash}X 
  | >{\centering\arraybackslash}X 
  | >{\centering\arraybackslash}X
  | >{\centering\arraybackslash}X
  | >{\centering\arraybackslash}X | }
\hline
$N$      & $Re_\lambda$ & L           & $\nu$ & $\epsilon$      & $l$                      \\ \hline
128      & 93           & 2$\upi$     & 0.005 & 0.324           & 1.48                     \\ \hline
$l/\eta$ & $u'$         & $u'/u_\eta$ & $T_L$ & $T_L/\tau_\eta$ & $\kappa_{{\rm max}}\eta$ \\ \hline
59.6     & 0.984        & 4.91        & 1.51  & 12.14           & 1.5                      \\ \hline
\end{tabularx}
	\caption{Simulation parameters for the DNS study of isotropic turbulence (arbitrary units).
		$N$ is the number of grid points in each direction, 
		$Re_\lambda \equiv u'\lambda/\nu$ is the Taylor micro-scale
		Reynolds number ($Re_\lambda \equiv \sqrt{15Re}$ for homogeneous and isotropic flows), $\lambda\equiv u'/\langle(\boldsymbol{\nabla} \boldsymbol{u})^2\rangle^{1/2} $ is the Taylor micro-scale,
		$\mathscr{L} $ is the box size, $\nu$ is the fluid kinematic viscosity, $\epsilon \equiv 2\nu \int_0^{\kappa_{\rm max}}\kappa^2 E(\kappa) {\rm d}\kappa $ is the mean
		turbulent kinetic energy dissipation rate,
		$l \equiv 3\upi/(2k)\int_0^{\kappa_{\rm max}}E(\kappa)/\kappa {\rm d}\kappa $  is the integral length scale, $\eta \equiv \nu^{3/4}/\epsilon^{1/4}$ is the Kolmogorov length scale, 
		$u' \equiv \sqrt{(2k/3)}$ is the fluid r.m.s. fluctuating 
		velocity, $k$ is the turbulent kinetic energy, 
		$u_\eta$ is the Kolmogorov velocity scale, 
		$T_L \equiv l/u^\prime$ is the large-eddy turnover
		time, $\tau_\eta \equiv \sqrt{(\nu/\epsilon)}$ is the Kolmogorov time scale, 
		$\kappa_{\rm max}=\sqrt{2}N/3$ is the maximum resolved wavenumber.}
	{\label{tab:parameters}}
\end{table}
\FloatBarrier

\subsection{Model Hyper-Parameters}

As explained in the previous section, the VQ-AE network provides a compressed representation of data in a discrete/quantized latent space by vector-quantizing the intermediate representations of the autoencoder. Such discrete representations are more compact than the original data, while the reconstructed data has minimal distortion. We design our network so that it can transform and downsample original data by a scaling factor of $SF \in \{2,4,8\}$ depending on the level of reconstruction quality and compression needed. With $K = 512$ representing the size of the codebook (number of codewords) and mapping three velocity components into one in the discrete latent space, we can achieve $\frac{3 \times 32}{1 \times 9} \times (SF)^{3}$ reduction in bits, corresponding to $85$, $683$ and $5461$ when $SF$ is $2$, $4$ and $8$, respectively. Indeed, an input data of shape $(3,128,128,128)$ is compressed to $(1,64,64,64)$ with $SF=2$, $(1,32,32,32)$ with $SF=4$ or $(1,16,16,16)$ with $SF=8$.

We trained three versions of the model in which all of the hyper-parameters (of frameworks) are the same except the scaling factor, $SF$. 
Throughout our framework, each covolution layer is followed by a batch normalization (BN) layer and a rectified linear unit activation function, so called $ReLU$ ($ReLU(x)=max(0,x)$). Embedding batch normalization layer  \cite{ioffe2015batch} adjusts (scales to unit variance and zero mean) the intermediate outputs of each convolutional layer so that their distributions are more stable throughout the framework. This batch normalization layer is frequently used in the deep learning community and is known to accelerate the training of deep neural nets.

\begin{figure}
	\centering	
	\includegraphics[width=1\linewidth]{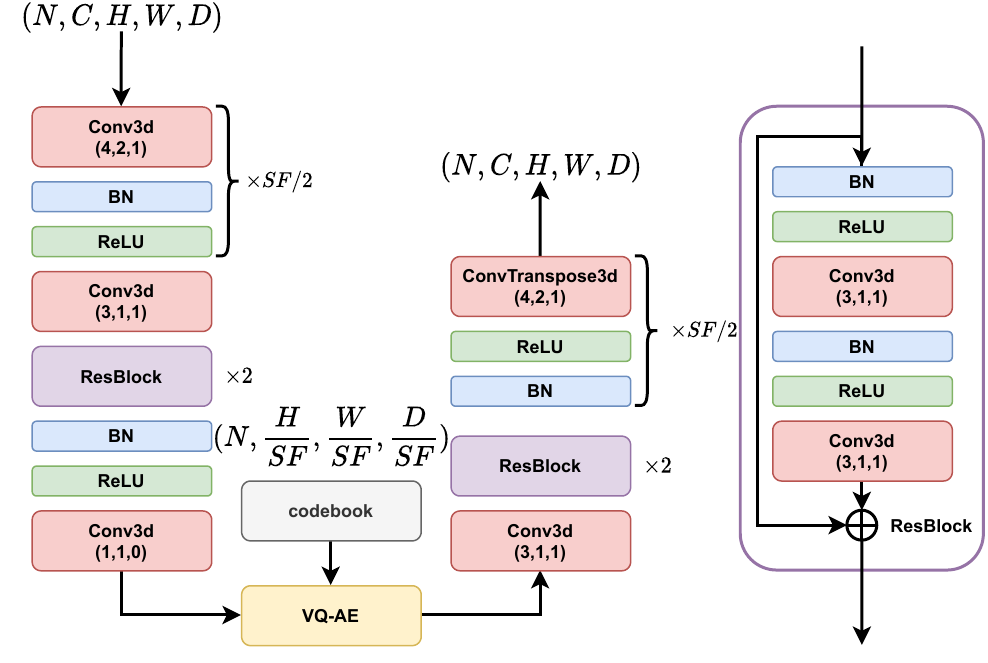}
	\caption{Schematic of the VQ-AE architecture. In this figure, $N$ is the number of training data in a batch of them, $C$ is the number of channels in the input data (corresponding with the number of velocity components), $H$ , $W$ and $D$ represent the height, width and depth of input data, and $SF \in \{2,4,8\}$ is the scaling factor.}
	\label{fig:VQ-AE_Schematic}
\end{figure}

We use three types of convolutional layers, type 1 with kernel size $4$ and stride length $2$ (used for the purpose of dimension reduction/expansion), type 2 with kernel size $3$ and stride length $1$ and type 3 with kernel size $1$ and stride length $1$ for learning transformation. The number of convolutional layers in their networks depends on the scaling factor $SF$. When $SF=2$, the encoder is composed of one type 1 kernel followed by a type 2 kernel, while the decoder consists only of one type 1 kernel. Similarly with $SF=8$, the encoder is composed of three type 1 kernels followed by one type 2 kernel, while the decoder has three type 1 kernels. We also use residual blocks (\cite{he2016deep}) to obtain a robust performance as the networks become deeper, without which there could arise optimization issues due to vanishing/exploding gradients. We embedded only 2 residual blocks, as opposed to 12 in \cite{glaws2020deep}, where each block is composed of two type 2 kernels convolutional layers which are embedded at the end of encoder and the beginning of decoder networks. Such a framework drastically reduces the number of learnable parameters compared to the recent fully convolutional autoencoder in \cite{glaws2020deep}. The schematic of this framework is shown in figure \ref{fig:VQ-AE_Schematic}.

\FloatBarrier

We designed a codebook with $K = 512$ codewords, each with embedding dimension $64$. For the loss function hyper-parameters, we use commitment loss coefficient $\beta = 0.25$ (as suggested in \cite{van2017neural}), $\alpha = 0.1$ and $a = 2$ for the velocity gradient constraint (motivated by the knowledge that at the small-scales of isotropic turbulence, the variance of the transverse velocity gradients are twice the size of the longitudinal ones \cite{pope_2000}), and finally $\gamma = 10^{-4}$ for the term corresponding to other constraints. It is worth mentioning that an improved reconstruction of the velocity gradient tensor would eventually lead to better results for the terms in the other constraint, which is the reason for setting $\alpha \gg \gamma$. We trained our framework for 200 epochs with batch size $=1$ using the Adam optimizer \cite{kingma2014adam} with learning rate = $0.001$, along with a learning rate scheduler. 
It should be noted that our focus has been on the general characteristics of the framework rather than achieving the best configurations, hence there might be room for improvements by tuning the proposed hyper-parameters. 
We implemented this framework in PyTorch using the CUDA environment, and trained it on one NVIDIA Pascal P100 GPU. The training was completed within approximately 8 hours, and the maximum GPU memory consumed was around 5 $GB$.

\vskip -0.5 cm
\section{Evaluation Metrics}\label{Evaluation_Metrics}
\vskip -0.5 cm

In addition to the conventional error measurement methods such as MSE and mean absolute error (MAE), the performance of the reconstructed field is assessed using additional diagnostic tests from the image processing community, as well as rigorous physics-based metrics relevant to the analysis of turbulence.
We employ two image similarity metrics (i) the peak signal-to-noise-ratio (PSNR) defined as

\begin{eqnarray}\label{eq:PSNR}
 PSNR (x,\hat{x}) = 10 \: log_{10} \Big(\frac{MAX(x,\hat{x})^{2}}{MSE(x,\hat{x})}\Big)
\end{eqnarray}
which, unlike the MSE, can handle grid data with different dynamic ranges,
and (ii) the mean structural similarity index measure (MSSIM) \cite{wang2004image}. For brevity, we do not repeat the concept or equations related to the MSSIM metric as it is already well explained in the recent study of \cite{glaws2020deep}. 

For the physics-based quality metrics, in addition to the already defined Betchov relations, we consider the turbulence kinetic energy spectra, and the probability density functions (PDFs) of the filtered velocity gradient tensor and its invariants. The turbulent kinetic energy spectra provides information on the kinetic energy content at different scales in the flow, providing crucial quantitative information on the flow. However, the energy spectrum does not provide information on the rich, multiscale statistical geometry in the flow which is crucial for the dynamics of turbulence, nor does it provide information on intermittent fluctuations in the flow. For example, the geometric alignments between the strain-rate and vorticity fields play an important role in the fundamental process of vortex stretching which in turn is crucial for the generation of intermittent fluctuations in the flow \cite{buaria20}, and these small scales processes in turbulence are captured by analyzing the velocity gradient tensor $\boldsymbol{A}$ \cite{tsinober,meneveau11}. Moreover, by filtering $\boldsymbol{A}$ at different scales to obtain the filtered velocity gradient tensor, $\widetilde{\boldsymbol{A}}$, one can analyze the multiscale properties of the flow and processes such as the energy cascade \cite{danish18,carbone20,johnson20,tom21}. Therefore, to provide a more thorough test of the quality of the reconstructed fields, in addition to the energy spectrum we also consider PDFs of the diagonal and off-diagonal components of $\widetilde{\boldsymbol{A}}$, as well as joint PDFs of the principal invariants of $\widetilde{\boldsymbol{A}}$, namely $Q\equiv-\frac{1}{2}tr(\widetilde{\boldsymbol{A}}^2)$ and $R\equiv-\frac{1}{2}tr(\widetilde{\boldsymbol{A}}^3)$. The joint PDF of $Q,R$ provides rich information and is commonly used to assess the basic properties of turbulent flows \cite{tsinober,meneveau11}, with $Q$ measuring the relative importance of strain and rotation rates at a point in the flow ($Q > 0$ indicates rotation dominated regions), while $R$ measures the relative importance of strain self-amplification (SSA) and vortex stretching (VS) at a point in the 
flow ($R > 0$ indicates SSA dominated region). In our results, to construct $\widetilde{\boldsymbol{A}}$ we convolve $\boldsymbol{A}$ with a low-pass Gaussian filter for three different filtering lengths, corresponding to filtering in the dissipation, inertial, and large scale (energy containing) ranges of the turbulent flow.

\vskip -0.5 cm

\section{Results and Discussion}\label{Results_Discussion}
\vskip -0.5 cm
\subsection{Statistically Stationary Isotropic Turbulence}
In Tables \ref{tab:Summary_HIT_SF2_Table}, \ref{tab:Summary_HIT_SF4_Table} and \ref{tab:Summary_HIT_SF8_Table} we summarize our results with respect to the performance of the trained VQ-AE model on an unseen realization from statistically stationary, isotropic turbulence.
The results for $SF=2$, corresponding to $CR=85$, clearly indicate that our model successfully passes all of the visual assessment and similarity-based metric tests. Moreover, the model predicts the second order statistics $\langle S_{ij} S_{ij} \rangle$ and $\langle R_{ij} R_{ij} \rangle$ with less than 5 percent error, and the third order statistics $\langle S_{ij} S_{jk} S_{ki}\rangle$ and $\langle S_{ij} \omega_{i} \omega_{j}\rangle$ with less than 10 percent error for the ``no filter'' (corresponding to the ``small scales'' column in the tables) case. This highlights that our model can capture up to third order statistics of the velocity gradient tensor with reasonable accuracy while offering $85 \times$ compression, meaning an $85 \times $ increase in data transfer and $85 \times $ decrease in disk usage. 

Our results also show that when the filter length is increased, the model results are even closer to the original, indicating that the loss of information during compression mainly took place at the very smallest scales of flow, which is a trend seen at all the three models with different $SF$. That ML-based compression methods lead to loss of information mainly at the smallest scales was also observed in \cite{glaws2020deep,mohan2020spatio}. For a given filter length, we also find that increasing the scale factor from $SF=2$ to $SF=4$ mainly affects the accuracy of the prediction of the third order statistics (and presumably also the higher-order statistics), while the second order statistics are still predicted to within 15 percent error at the small scales (i.e. for the no-filter case), and the quality of the results at the inertial and large scales of the flow are almost the same as those for $SF=2$. Therefore, our model with $SF=4$ (corresponding to $CR=683$) captures most of the information content of the flow at the inertial scales of flow, and may be employed for the situations where there is less interest in accurate representations of the smallest scales of the flow (e.g. they may be of less interest in some cases since they contain relatively little kinetic energy). Similarly, our model with $SF=8$ (corresponding to $CR=5461$) may be used when the main interest is in accurately representing the large scales of flow. 

For the rest of this paper, we present only the results for the $SF=2$ case. Furthermore, while not presented here, for each choice of $SF$, our model satisfies the incompressibility constraint on the flow well, with $\max[A_{ii}] = O(10^{-10})$.

\begin{table}[]
\centering
\begin{tabular}{cccc}
\hline
Metrics                                                                                                                & (SF=2, Small scales) & ($SF=2$,Inertialscales) & ($SF=2$, Large scales) \\ \hline
MSE                                                                                                                     & $4.39\times10^{-3}$    & $3.53\times10^{-3}$       & $2.84\times10^{-3}$  \\
MAE                                                                                                                    & $4.99\times10^{-2}$    & $4.54\times10^{-2}$       & $4.12\times10^{-2}$   \\
PSNR                                                                                                                   & $34.98$                & $35.45$                   & $35.89$               \\
MSSIM                                                                                                                  & $0.977$                & $0.980$                   & $0.982$               \\
$\langle S_{ij} S_{ij} \rangle$,$\langle \widehat{ S_{ij}   S_{ij}} \rangle$                                           & $33.02\:,\:31.32$    & $19.82\:,\:18.87$     & $10.33\:,\:9.78$     \\
$\langle R_{ij} R_{ij} \rangle$,$\langle \widehat{ R_{ij}   R_{ij}} \rangle$                                           & $33.03\:,\:31.06$    & $19.82\:,\:18.83$     & $10.33\:,\:9.77$     \\
$(-3/4)\times$ ($\langle S_{ij} \omega_{j} \omega_{j}\rangle$,$\langle \widehat{S_{ij} \omega_{j} \omega_{j}}\rangle$) & $-61.76\:,\:-54.19$   & $-26.05\:,\:-23.84$   & $-9.26\:,\:-8.44$    \\
$\langle S_{ij} S_{kj} S_{ji} \rangle$,$\langle \widehat{ S_{ij} S_{kj} S_{ji}} \rangle$                               & $-61.76\:,\:-54.63$  & $-26.05\:,\:-24.03$   & $-9.26\:,\:-8.49$    \\ \hline
\end{tabular}
\caption{Summary of the performance of trained VQ-AE evaluated on an unseen data from statistically stationary isotropic turbulence. $SF=2$ represents scaling the input data ($3 \times 128^{3}$) by a factor of two which yields a compressed integer representation with size $1 \times 64^{3}$ and $CR=85$.}
\label{tab:Summary_HIT_SF2_Table}
\end{table}

\begin{table}[]
\centering
\begin{tabular}{cccc}
\hline
Metrics                                                                                                                & ($SF=4$, Small scales) & ($SF=4$, Inertialscales) & ($SF=4$, Large scales) \\ \hline
MSE                                                                                                                    & $2.01\times10^{-2}$    & $1.56\times10^{-2}$       & $1.21\times10^{-2}$    \\
MAE                                                                                                                    & $1.07\times10^{-1}$    & $9.45\times10^{-2}$       & $8.34\times10^{-2}$    \\
PSNR                                                                                                                   & $28.36$                & $28.99$                   & $25.59$                 \\
MSSIM                                                                                                                  & $0.909$                & $0.923$                   & $0.934$                \\
$\langle S_{ij} S_{ij} \rangle$,$\langle \widehat{ S_{ij} S_{ij}} \rangle$                                             & $33.02\:,\: 28.17$          & $19.82\:,\: 17.75$             & $10.33\:,\: 9.32$        \\
$\langle R_{ij} R_{ij} \rangle$,$\langle \widehat{ R_{ij} R_{ij}} \rangle$                                             & $33.03\:,\: 27.35$          & $19.82\:,\: 17.56$             & $10.33\:,\: 9.25$  \\
$(-3/4)\times$ ($\langle S_{ij} \omega_{j} \omega_{j}\rangle$,$\langle \widehat{S_{ij} \omega_{j} \omega_{j}}\rangle$) &  $-61.76\:,\: -41.57$        & $-26.05\:,\:-20.98$           & $-9.26\:,\: -7.57$        \\
$\langle  S_{ij} S_{kj} S_{ji} \rangle$,$\langle \widehat{ S_{ij} S_{kj} S_{ji}} \rangle$                              & $-61.76\:,\:-41.23$        & $-26.05\:,\:-21.35$           & $-9.26\:,\:-7.73$  \\ \hline
\end{tabular}
\caption{Summary of the performance of trained VQ-AE evaluated on an unseen data from statistically stationary isotropic turbulence. $SF=4$ represents scaling the input data ($3 \times 128^{3}$) by a factor of four which yields a compressed integer representation with size $1 \times 32^{3}$ and $CR=683$.}
\label{tab:Summary_HIT_SF4_Table}
\end{table}

\begin{table}[]
\centering
\begin{tabular}{cccc}
\hline
Metrics                                                                                                                & ($SF=8$, Small scales) & ($SF=8$, Inertialscales) & ($SF=8$, Large scales) \\ \hline
MSE                                                                                                                    & $1.90\times10^{-1}$    & $1.67\times10^{-1}$      & $1.44\times10^{-1}$    \\
MAE                                                                                                                    & $3.24\times10^{-1}$    & $3.0\times10^{-1}$       & $2.74\times10^{-1}$    \\
PSNR                                                                                                                   & $18.61$                & $18.71$                  & $18.84$                \\
MSSIM                                                                                                                  & $0.6$                  & $0.645$                  & $0.686$                \\
$\langle S_{ij} S_{ij} \rangle$,$\langle \widehat{ S_{ij} S_{ij}} \rangle$                                             & $33.02\:,\:19.81$      & $19.82\:,\:13.16$        & $10.33\:,\:7.52$       \\
$\langle R_{ij} R_{ij} \rangle$,$\langle \widehat{ R_{ij} R_{ij}} \rangle$                                             & $33.03\:,\:17.43$      & $19.82\:,\:12.48$        & $10.33\:,\:7.20$       \\
$(-3/4)\times$ ($\langle S_{ij} \omega_{j} \omega_{j}\rangle$,$\langle \widehat{S_{ij} \omega_{j} \omega_{j}}\rangle$) & $-61.76\:,\:-18.3$     & $-26.05\:,\:-12.14$      & $-9.26\:,\:-5.62$      \\
$\langle  S_{ij} S_{kj} S_{ji} \rangle$,$\langle \widehat{ S_{ij} S_{kj} S_{ji}} \rangle$                              & $-61.76\:,\:-21.44$    & $-26.05\:,\:-13.06$      & $-9.26\:,\:-5.62$      \\ \hline
\end{tabular}\caption{Summary of the performance of trained VQ-AE evaluated on an unseen data from statistically stationary isotropic turbulence. $SF=8$ represents scaling the input data ($3 \times 128^{3}$) by a factor of eight which yields a compressed integer representation with size $1 \times 16^{3}$ and $CR=5461$.}{\label{tab:Summary_HIT_SF8_Table}}

\end{table}

\begin{figure}[h]
	\centering	
	\includegraphics[width=0.7\linewidth]{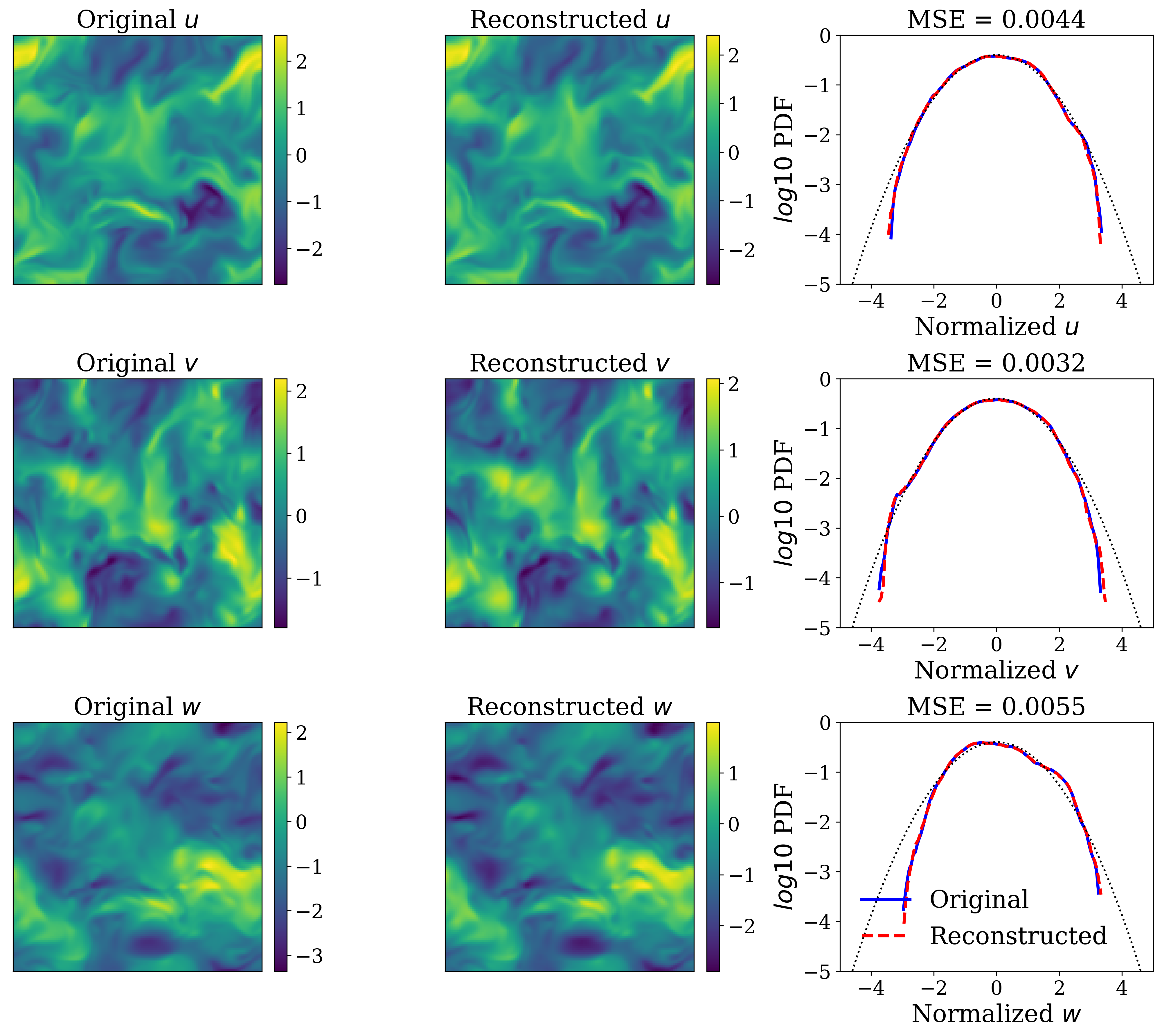}
	\caption{Comparing original and reconstructed 3D HIT compressed by VQ-AE with $SF=2$ (compression ratio  $CR = 85$): 2D snapshots and PDFs of velocity components}
	\label{fig:HIT_uvw_0_Turb_uvw_vqvae_1_exact}
\end{figure}

In figure \ref{fig:HIT_uvw_0_Turb_uvw_vqvae_1_exact}, we evaluate the performance of our ($SF=2$) model in reconstructing 2d snapshots (which are sampled randomly) of the velocity field, as well as the PDFs of the velocity components (where the statistics are based on a single snapshot, averaged over the full 3d domain). The snapshot comparisons show that our model captures very well the instantaneous spatial structure of the flow, and the PDF results show that the model accurately captures the statistical properties of the velocity field

\FloatBarrier

\begin{figure}[h]
	\centering	
	\includegraphics[width=0.4\linewidth]{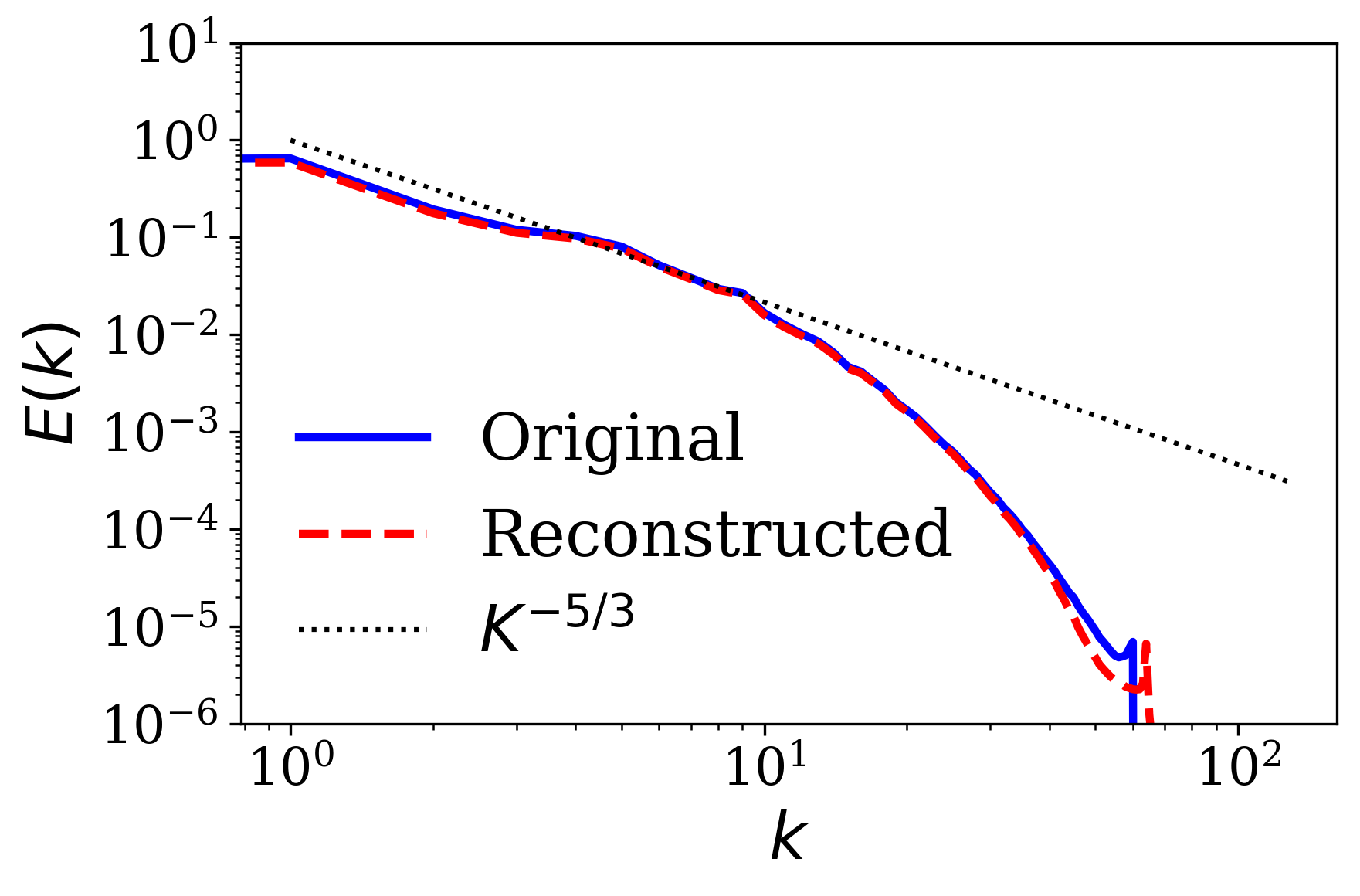}
	\caption{Comparing original and reconstructed 3D HIT compressed by VQ-AE with $SF=2$ (compression ratio  $CR = 85$): Turbulence Kinetic Energy Spectra.}
	\label{fig:HIT_EnergySpectrum_0_Turb_uvw_vqvae_1_exact}
\end{figure}

As already seen based on the results in the table \ref{tab:Summary_HIT_SF2_Table}, our model accurately captures the large and inertial scales of flow, with some loss of information for the very smallest scales. This can also be seen in the TKE spectra shown in figure \ref{fig:HIT_EnergySpectrum_0_Turb_uvw_vqvae_1_exact}. Our results indicate a fascinating improvement compared to the recent study of \cite{mohan2020spatio}, where their fully CNN AE model with $CR = 125$ captured the large scales completely, and the inertial scales with some distortion, but was in significant error for small scales, both quantitatively and qualitatively.

\FloatBarrier

The PDFs of the longitudinal and transverse components of the velocity gradient tensor are shown in in figure \ref{fig:HIT_VG_Turb_uvw_vqvae_1_exact} for different filtering lengths. The results illustrate that our model accurately captures the shape of these PDFs, including their heavy tails and skewness, with little distortion even for the very intermittent contributions that govern the tails of these PDFs. These results from our model show remarkable improvement compared to those from the recent study of \cite{mohan2020spatio}, whose model failed significantly in capturing the tails of these PDFs.

\begin{figure}
	\centering
	\begin{subfigure}[b]{\linewidth}
		\includegraphics[width=0.85\linewidth]{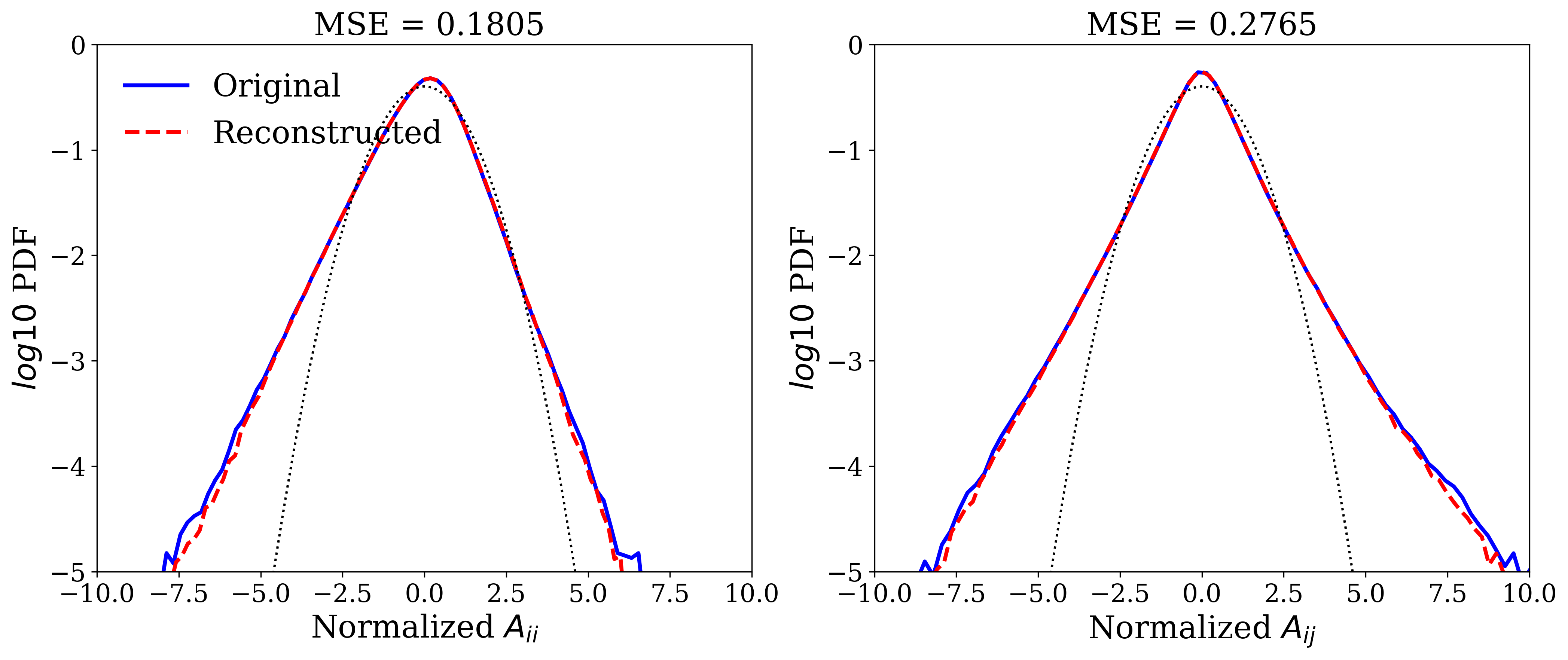}
		\caption{No filter}
	\end{subfigure}
	\begin{subfigure}[b]{\linewidth}
		\includegraphics[width=0.85\linewidth]{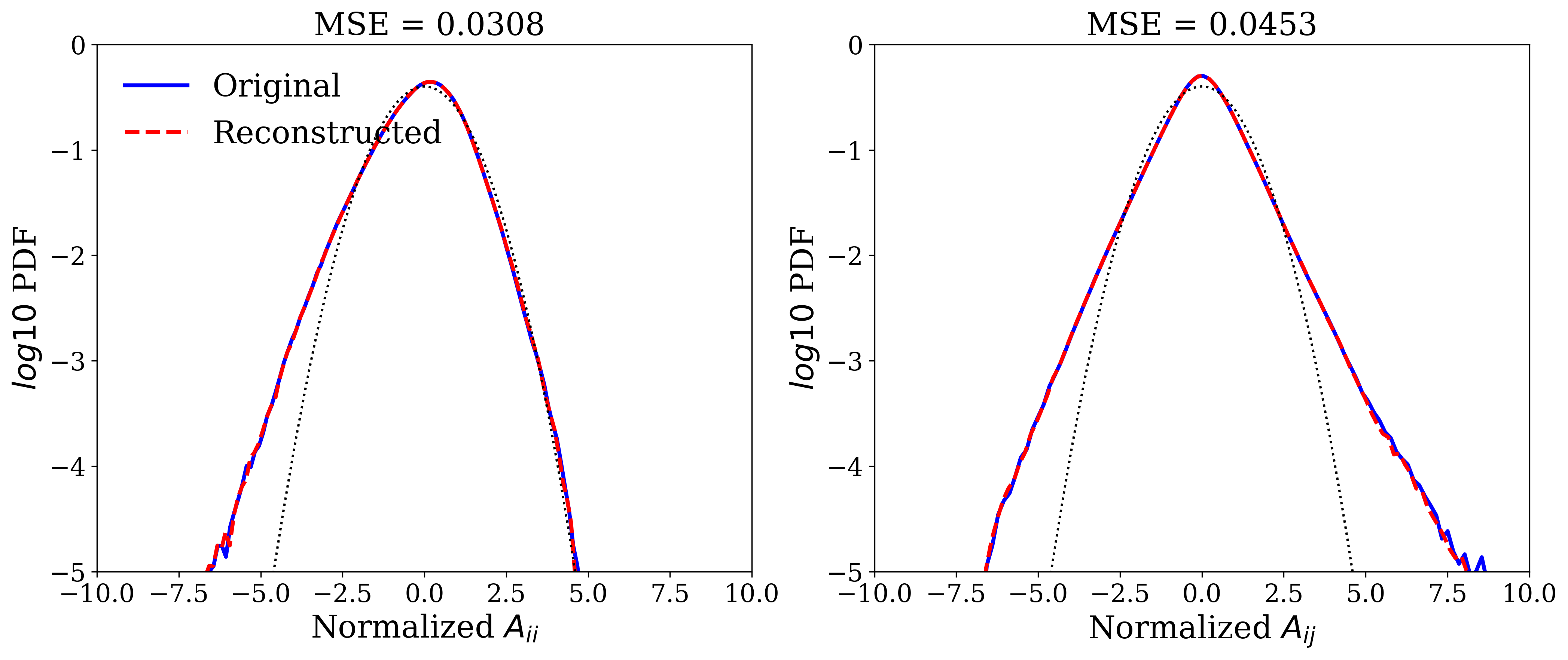}
		\caption{Inertial scales}
	\end{subfigure}
		\begin{subfigure}[b]{\linewidth}
		\includegraphics[width=0.85\linewidth]{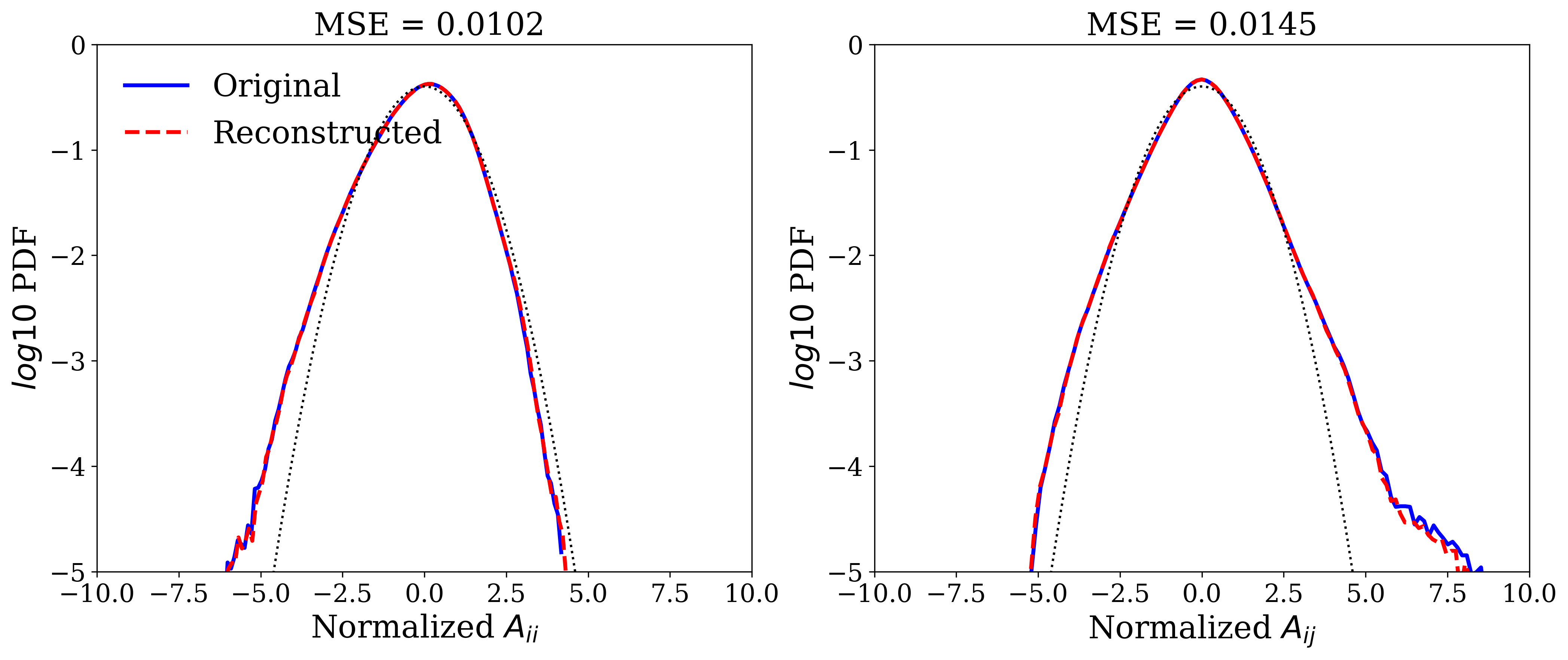}
		\caption{Large scales}
	\end{subfigure}
	\caption{Comparing original and reconstructed 3D HIT compressed by VQ-AE with $SF=2$ (compression ratio  $CR = 85$): PDFs of normalized longitudinal (on the left side) and transverse (on the right side) components of velocity gradient tensor $\boldsymbol{A}$.}\label{fig:HIT_VG_Turb_uvw_vqvae_1_exact}
\end{figure}
\FloatBarrier

\begin{figure}[h]
	\centering
		\includegraphics[width=\linewidth]{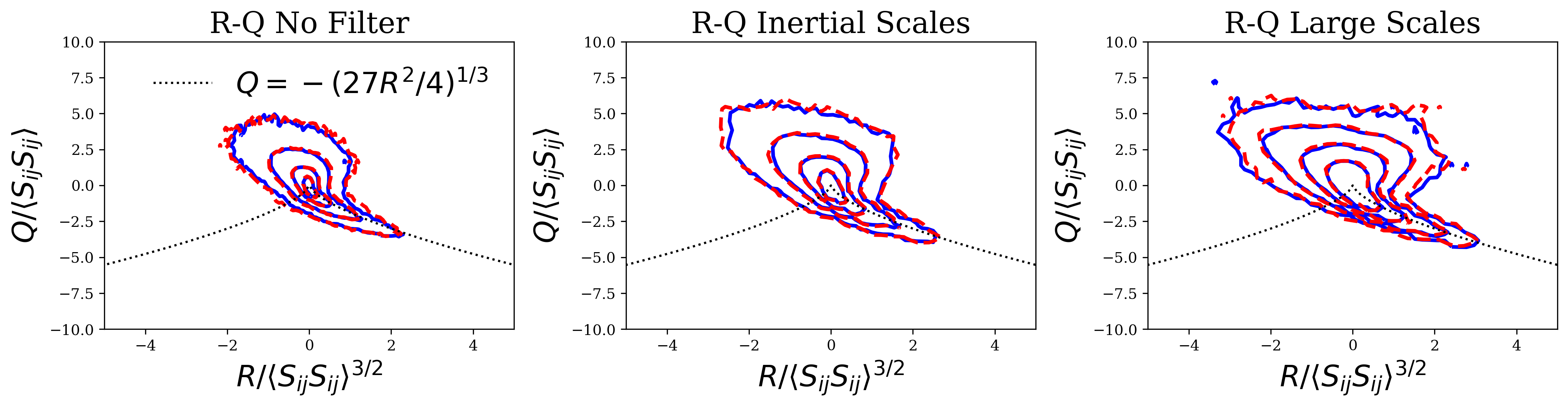}
	\caption{Comparing original and reconstructed 3D HIT compressed by VQ-AE with $SF=2$ (compression ratio  $CR = 85$): Contour plots of the joint PDF of normalized $Q=-Tr(\boldsymbol{A}^{2})/2$ and $R=-Tr(\boldsymbol{A}^{3})/3$.}\label{fig:HIT_RQ_Turb_uvw_vqvae_1_exact}
\end{figure}

The models capability in reconstructing the joint-PDF of the $Q$ and $R$ invariants of velocity gradient tensor is shown in figure \ref{fig:HIT_RQ_Turb_uvw_vqvae_1_exact}. The model accurately captures the distinctive ``tear-drop'' shape \cite{meneveau11,tsinober} of the PDF, significantly outperforming the model in \cite{mohan2020spatio}.

\FloatBarrier

\subsection{Effect of Regularization term}

\begin{figure}[h]
	\centering
	\begin{subfigure}[b]{0.5\linewidth}
		\includegraphics[width=\linewidth]{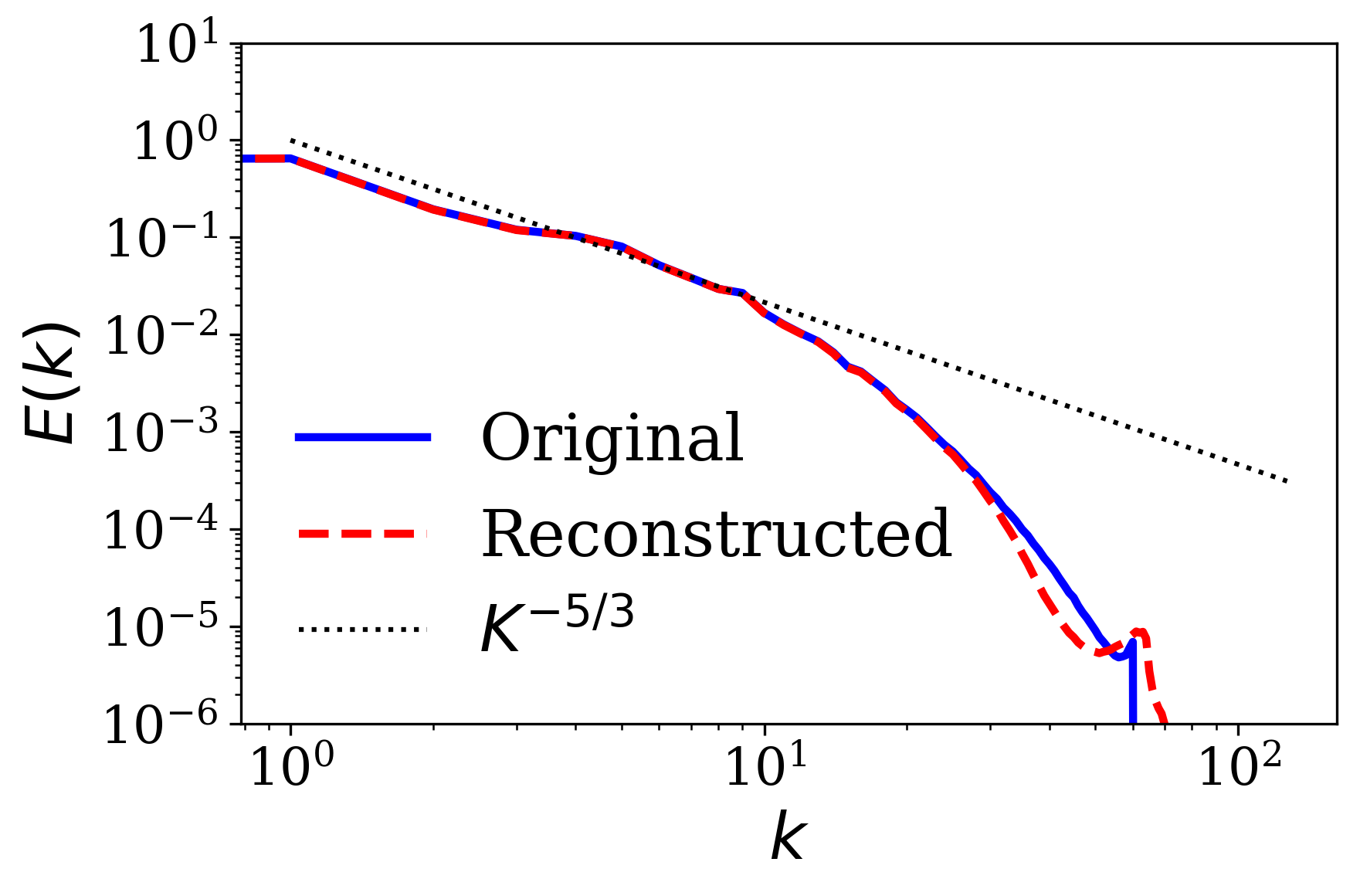}
		\caption{No Regularization}
	\end{subfigure}%
	\begin{subfigure}[b]{0.5\linewidth}
		\includegraphics[width=\linewidth]{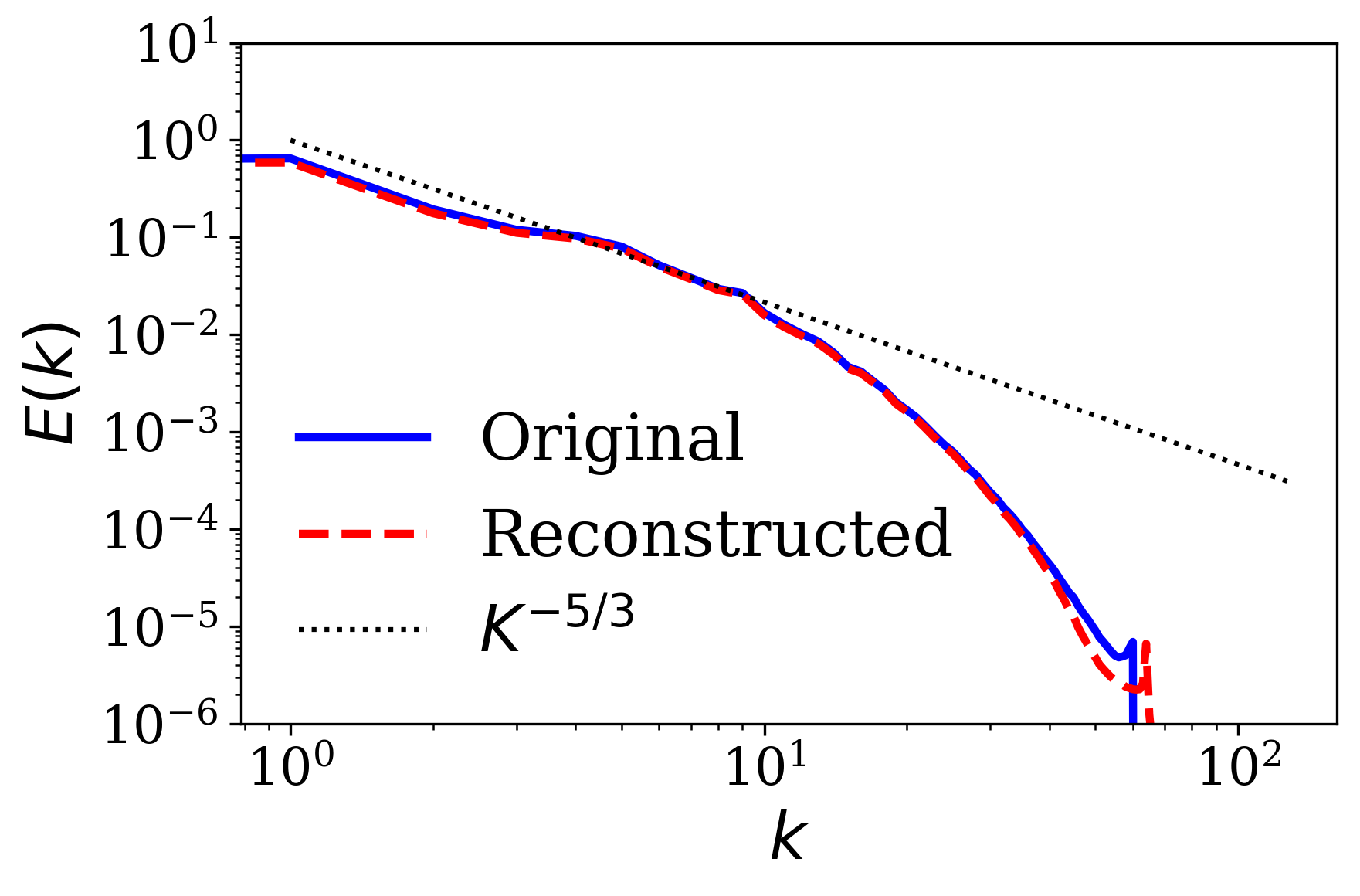}
		\caption{with Regularization}
	\end{subfigure}
	\caption{Comparing original and reconstructed 3D HIT compressed by VQ-AE with $SF=2$ (compression ratio  $CR = 85$). }\label{fig:HIT_Regularization_EnergySpectrum_Turb_uvw_vqvae_1_exact}
\end{figure}

To better understand the effect of calibrating the loss function by incorporating prior knowledge of the properties of the data, we trained another model without those regularization terms, which amounts to setting $\alpha = \gamma =0$ in the model. As shown in figure \ref{fig:HIT_Regularization_EnergySpectrum_Turb_uvw_vqvae_1_exact}, including the regularization term makes some improvement to the energy spectrum predictions at the smallest scales (highest wavenumbers). The results in figure \ref{fig:HIT_VG_Turb_uvw_vqvae_1_exact} show that including the regularization term also enhances the ability of the model to capture the intermittent fluctuations of the velocity gradient, characterized by the tails of the PDFs. Nevertheless, While there is improvement, the model without regularization already does a very good job at capturing the properties of the turbulent flow, and that for other flow quantities that we considered, the difference in the predictions from the model with and withour regularization is minimal.

\FloatBarrier

\begin{figure}[h]
	\centering
	\begin{subfigure}[b]{\linewidth}
		\includegraphics[width=0.9\linewidth]{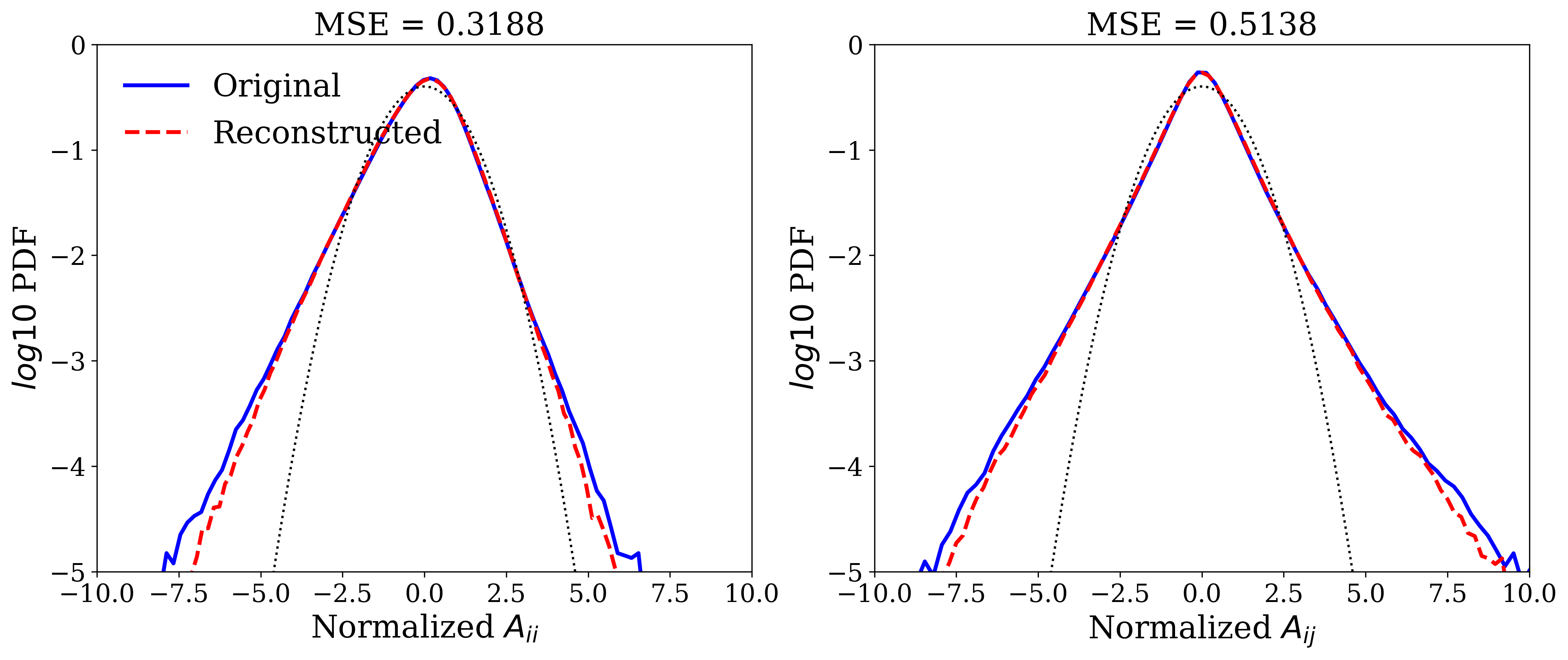}
		\caption{No Regularization}
	\end{subfigure}
	\begin{subfigure}[b]{\linewidth}
		\includegraphics[width=0.9\linewidth]{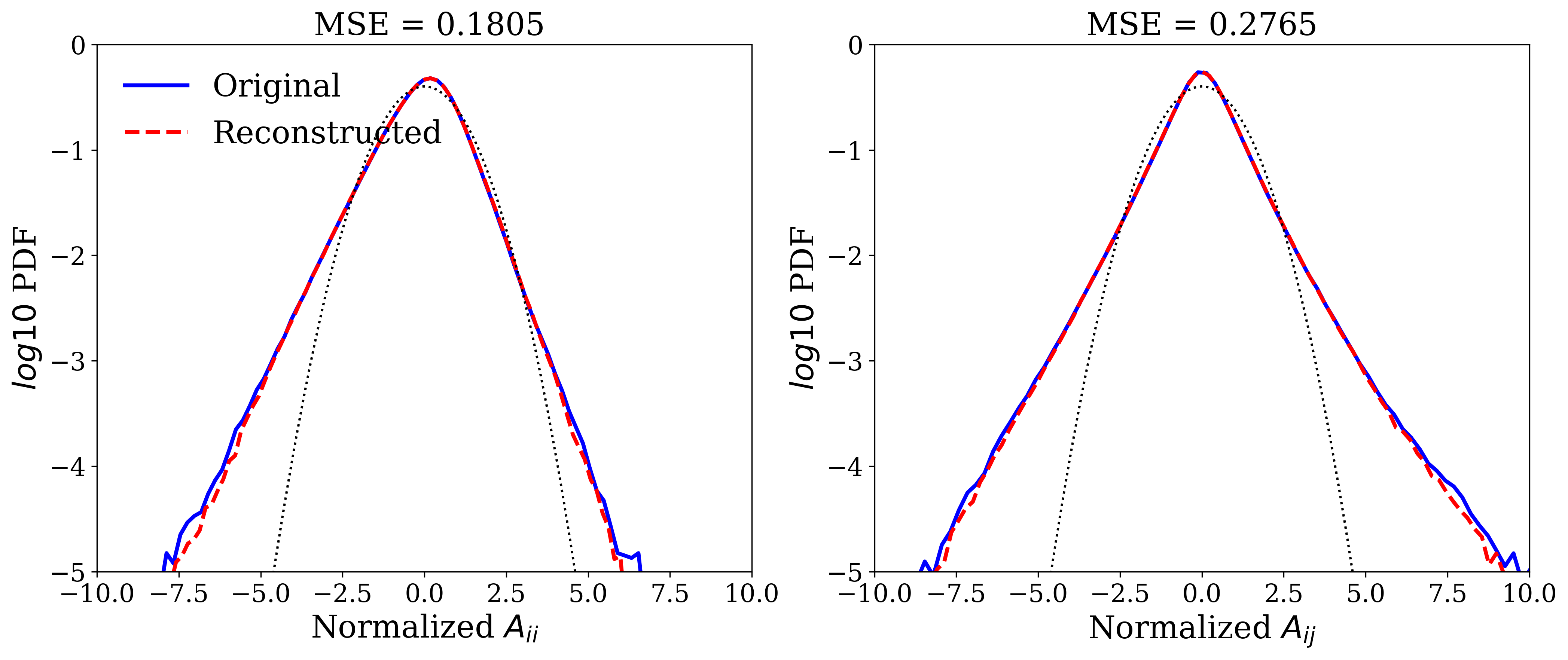}
		\caption{with Regularization}
	\end{subfigure}
	\caption{Comparing original and reconstructed 3D HIT compressed by VQ-AE with $SF=2$ (compression ratio  $CR = 85$): PDFs of normalized longitudinal (diagonal) and transverse (off-diagonal) components of velocity gradient tensor $\boldsymbol{A}$. }\label{fig:HIT_VG_Turb_uvw_vqvae_1_exact}
\end{figure}
\FloatBarrier

\subsection{Decaying Isotropic Turbulence}
In Tables \ref{tab:Summary_DIT_SF2_Table}, \ref{tab:Summary_DIT_SF4_Table} and \ref{tab:Summary_DIT_SF8_Table} we summarize our results with respect to the performance of the trained VQ-AE on a realization from decaying isotropic turbulence (DIT), which is a different flow from that which the model was trained on. We used the same flow snapshot of DIT as that used in the recent study of \cite{glaws2020deep}, allowing us to make a fair comparison between our model results and theirs.

Compared to the results presented in \cite{glaws2020deep}, where their fully convolutional AE model provided $CR=64$, our $SF=2$ model improves both the $MSE$ and $MAE$ by an order of magnitude, the $PSNR$ by 10\%, and the $MSSIM$ by 2\%, while offering a $CR=85$ which corresponds to more than a 30\% enhancement in the compressive capabilities.

\begin{table}[]
\centering
\begin{tabular}{cccc}
\hline
Metrics                                                                                                                & ($SF=2$, Small scales) & ($SF=2$, Inertialscales) & ($SF=2$, Large scales) \\ \hline
MSE                                                                                                                    & $1.84\times10^{-3}$    & $1.28\times10^{-3}$      & $9.67\times10^{-4}$    \\
MAE                                                                                                                    & $3.26\times10^{-2}$    & $2.73\times10^{-2}$      & $2.38\times10^{-2}$    \\
PSNR                                                                                                                   & $34.26$                & $34.34$                  & $34.14$                \\
MSSIM                                                                                                                  & $0.970$                & $0.974$                  & $0.975$                \\
$\langle S_{ij} S_{ij} \rangle$,$\langle \widehat{ S_{ij} S_{ij}} \rangle$                                             & $16.27\:,\:14.82$      & $8.93\:,\:8.54$          & $4.62\:,\:4.42$        \\
$\langle R_{ij} R_{ij} \rangle$,$\langle \widehat{ R_{ij} R_{ij}} \rangle$                                             & $16.27\:,\:14.70$      & $8.93\:,\:8.53$          & $4.62\:,\:4.42$        \\
$(-3/4)\times$ ($\langle S_{ij} \omega_{j} \omega_{j}\rangle$,$\langle \widehat{S_{ij} \omega_{j} \omega_{j}}\rangle$) & $-21.58\:,\:-18.09$    & $-8.11\:,\:-7.48$        & $-2.68\:,\:-2.49$      \\
$\langle  S_{ij} S_{kj} S_{ji} \rangle$,$\langle \widehat{ S_{ij} S_{kj} S_{ji}} \rangle$                              & $-21.58\:,\:-18.28$    & $-8.11\:,\:-7.57$        & $-2.68\:,\:-2.51$      \\ \hline
\end{tabular}	\caption{Summary of the performance of trained VQ-AE evaluated on an unseen data from decaying stationary isotropic turbulence. $SF=2$ represents scaling the input data ($3 \times 128^{3}$) by a factor of two which yields a compressed integer representation with size $1 \times 64^{3}$ and $CR=85$.}{\label{tab:Summary_DIT_SF2_Table}}
\end{table}

\begin{table}[]
\centering
\begin{tabular}{cccc}
\hline
Metrics                                                                                                                & ($SF=4$, Small scales) & ($SF=4$, Inertialscales) & ($SF=4$, Large scales) \\ \hline
MSE                                                                                                                    & $8.04\times10^{-3}$    & $5.40\times10^{-3}$      & $3.71\times10^{-3}$    \\
MAE                                                                                                                    & $6.93\times10^{-2}$    & $5.74\times10^{-2}$      & $4.79\times10^{-2}$    \\
PSNR                                                                                                                   & $27.85$                & $28.1$                   & $28.3$                 \\
MSSIM                                                                                                                  & $0.882$                & $0.899$                  & $0.909$                \\
$\langle S_{ij} S_{ij} \rangle$,$\langle \widehat{ S_{ij} S_{ij}} \rangle$                                             & $16.27\: ,   \:13.74$  & $8.93\:,\:8.47$          & $4.62\:,\:4.48$        \\
$\langle R_{ij} R_{ij} \rangle$,$\langle \widehat{ R_{ij} R_{ij}} \rangle$                                             & $16.27\:,\:13.27$      & $8.93\:,\:8.38$          & $4.62\:,\:4.45$        \\
$(-3/4)\times$ ($\langle S_{ij} \omega_{j} \omega_{j}\rangle$,$\langle \widehat{S_{ij} \omega_{j} \omega_{j}}\rangle$) & $-21.58\:,\:-14.31$    & $-8.11\:,\:-7.05$        & $-2.68\:,\:-2.41$      \\
$\langle  S_{ij} S_{kj} S_{ji} \rangle$,$\langle \widehat{ S_{ij} S_{kj} S_{ji}} \rangle$                              & $-21.58\:,\:-13.99$    & $-8.11\:,\:-7.14$        & $-2.68\:,\:-2.46$      \\ \hline
\end{tabular}	\caption{Summary of the performance of trained VQ-AE evaluated on an unseen data from decaying stationary isotropic turbulence. $SF=4$ represents scaling the input data ($3 \times 128^{3}$) by a factor of four which yields a compressed integer representation with size $1 \times 32^{3}$ and $CR=683$.}{\label{tab:Summary_DIT_SF4_Table}}
\end{table}

\begin{table}[]
\centering
\begin{tabular}{cccc}
\hline
Metrics                                                                                                                & ($SF=8$, Small scales) & ($SF=8$, Inertialscales) & ($SF=8$, Large scales) \\ \hline
MSE                                                                                                                    & $5.04\times10^{-2}$    & $3.85\times10^{-2}$      & $2.86\times10^{-2}$    \\
MAE                                                                                                                    & $1.72\times10^{-1}$    & $1.51\times10^{-1}$      & $1.3\times10^{-1}$     \\
PSNR                                                                                                                   & $19.88$                & $19.52$                  & $19.44$                \\
MSSIM                                                                                                                  & $0.598$                & $0.644$                  & $0.685$                \\
$\langle S_{ij} S_{ij} \rangle$,$\langle \widehat{ S_{ij} S_{ij}} \rangle$                                             & $16.28\:,\:10.58$      & $8.93\:,\:6.74$          & $4.62\:,\:4.08$        \\
$\langle R_{ij} R_{ij} \rangle$,$\langle \widehat{ R_{ij} R_{ij}} \rangle$                                             & $16.27\:,\:8.88$       & $8.93\:,\:6.33$          & $4.62\:,\:3.89$        \\
$(-3/4)\times$ ($\langle S_{ij} \omega_{j} \omega_{j}\rangle$,$\langle \widehat{S_{ij} \omega_{j} \omega_{j}}\rangle$) & $-21.58\:,\:-6.66$     & $-8.11\:,\:-4.54$        & $-2.68\:,\:-2.03$      \\
$\langle  S_{ij} S_{kj} S_{ji} \rangle$,$\langle \widehat{ S_{ij} S_{kj} S_{ji}} \rangle$                              & $-21.58\:,\:-11.76$    & $-8.11\:,\:-5.22$        & $-2.68\:,\:-2.21$      \\ \hline
\end{tabular}	\caption{Summary of the performance of trained VQ-AE evaluated on an unseen data from decaying stationary isotropic turbulence. $SF=8$ represents scaling the input data ($3 \times 128^{3}$) by a factor of eight which yields a compressed integer representation with size $1 \times 16^{3}$ and $CR=5461$.}{\label{tab:Summary_DIT_SF8_Table}}
\end{table}

\begin{figure}[h]
	\centering	
	\includegraphics[width=0.7\linewidth]{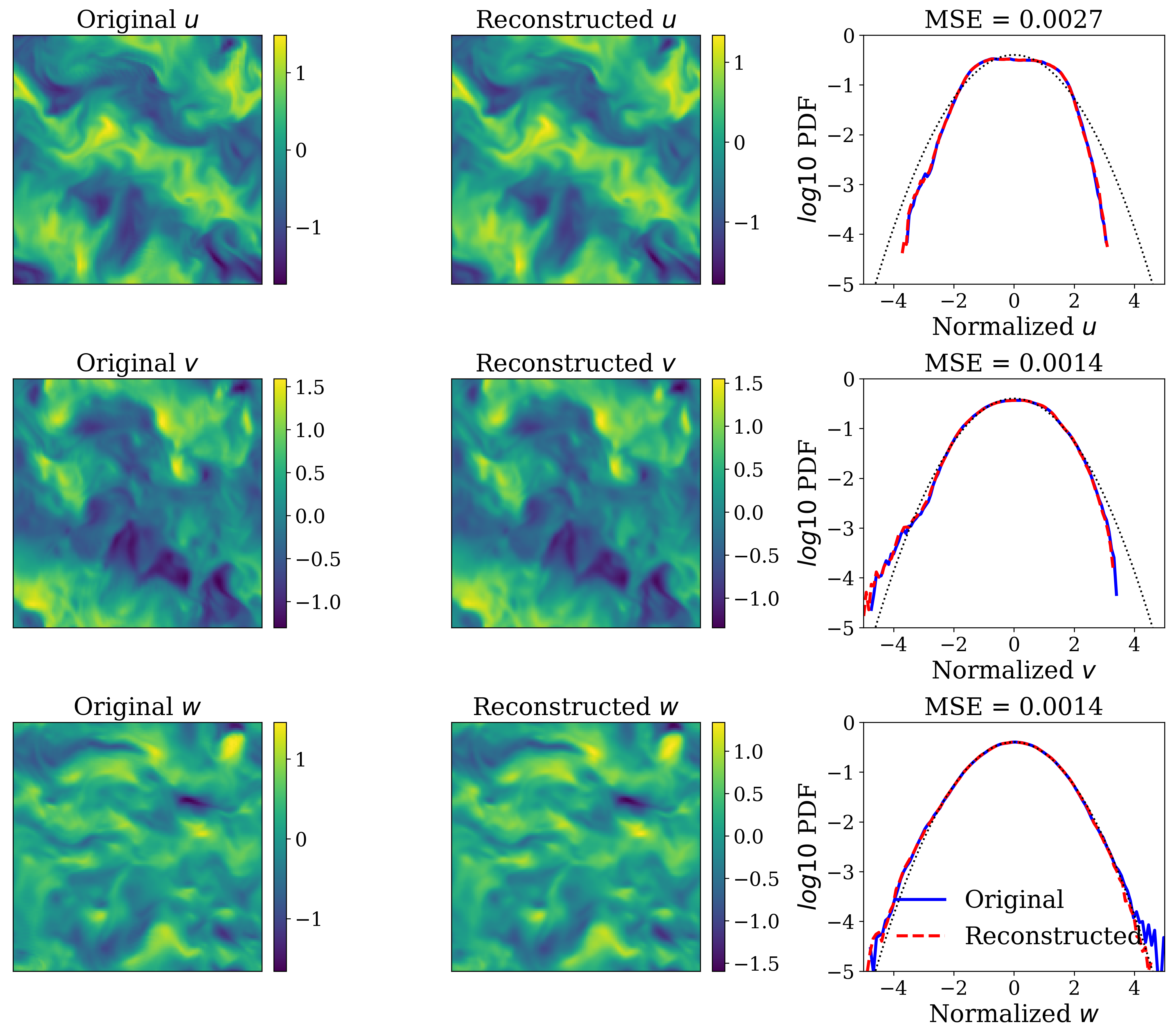}
	\caption{Comparing original and reconstructed 3D DIT compressed by VQ-AE with $SF=2$ (compression ratio  $CR = 85$): 2D snapshots and PDFs of velocity components}
	\label{fig:DIT_uvw_0_Turb_uvw_vqvae_1_exact}
\end{figure}

A visual comparison of the original and reconstructed velocity fields in DIT is shown in figure \ref{fig:DIT_uvw_0_Turb_uvw_vqvae_1_exact}. The results illustrate the remarkable performance of our model in capturing flow characteristics not seen during the training process.

\FloatBarrier

\begin{figure}[h]
	\centering	
	\includegraphics[width=0.4\linewidth]{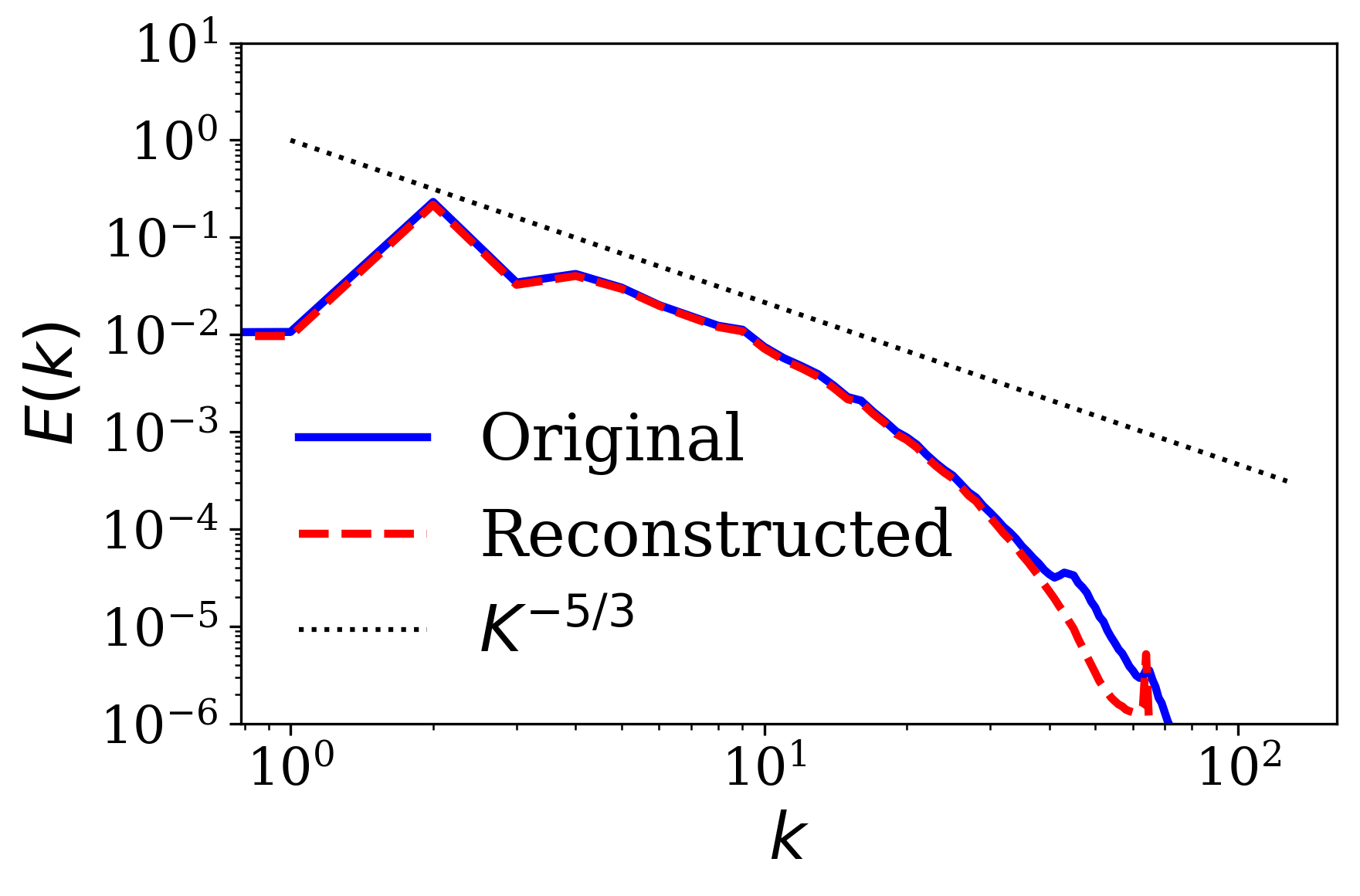}
	\caption{Comparing original and reconstructed 3D DIT compressed by VQ-AE with $SF=2$ (compression ratio  $CR = 85$): Turbulence Kinetic Energy Spectra }
	\label{fig:DIT_EnergySpectrum_0_Turb_uvw_vqvae_1_exact}
\end{figure}

In figure \ref{fig:DIT_EnergySpectrum_0_Turb_uvw_vqvae_1_exact} we show the results for the TKE spectrum, which illustrate the ability of our VQ-AE model to accurately capture the large and inertial scales of DIT, but with some deviation at the smallest scales. Compared to the model reconstruction result in \cite{glaws2020deep}, our model performs significantly better, being able to accurately capture the behavior over a wider range of scales.

\FloatBarrier
%

\begin{figure}[h]
	\centering
	\begin{subfigure}[b]{\linewidth}
		\includegraphics[width=0.85\linewidth]{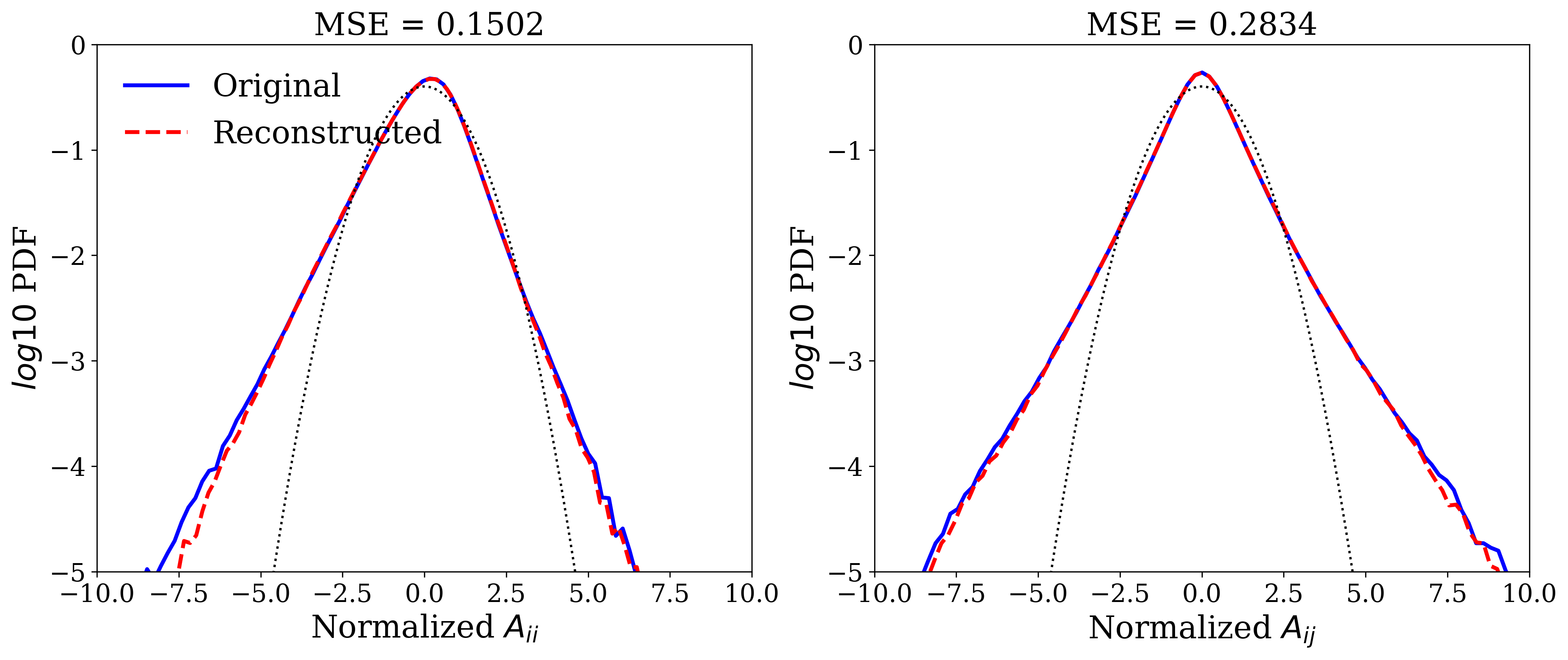}
		\caption{No filter}
	\end{subfigure}
	\begin{subfigure}[b]{\linewidth}
		\includegraphics[width=0.85\linewidth]{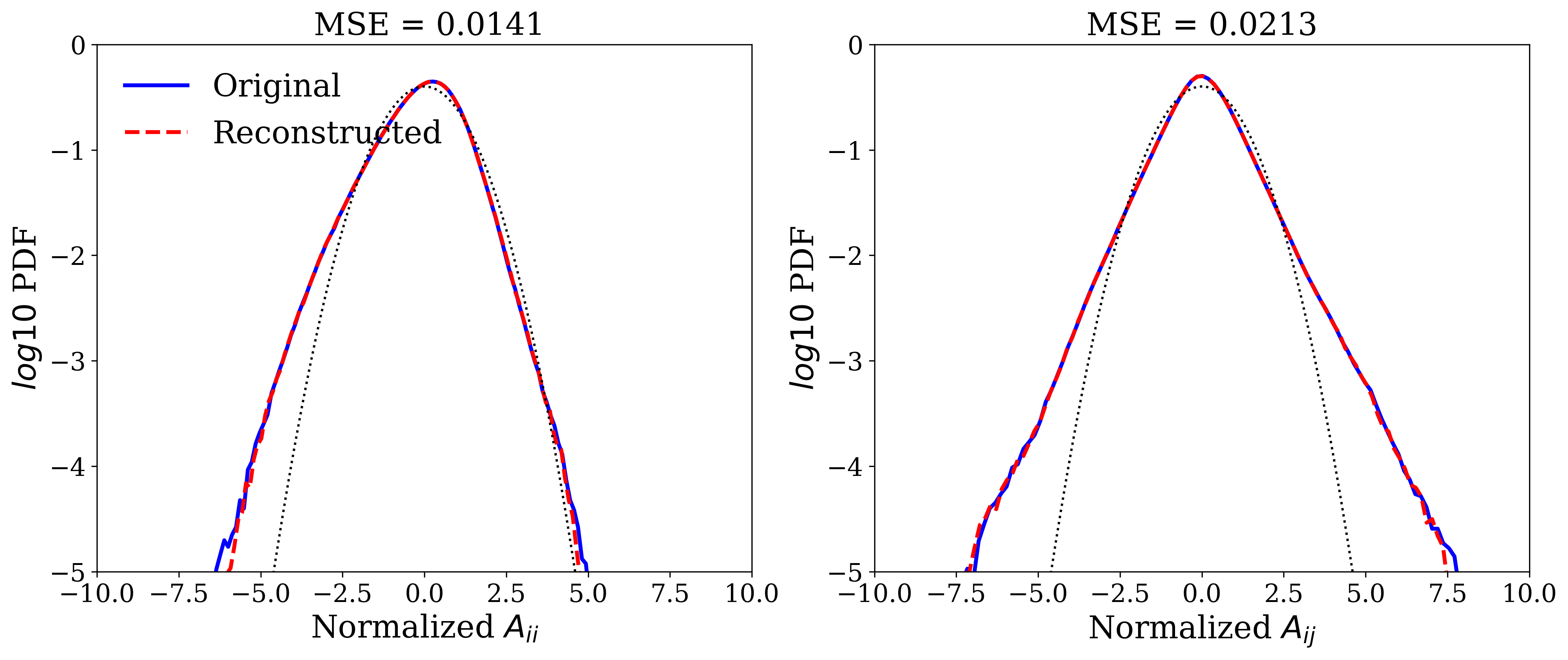}
		\caption{Inertial scales}
	\end{subfigure}
		\begin{subfigure}[b]{\linewidth}
		\includegraphics[width=0.85\linewidth]{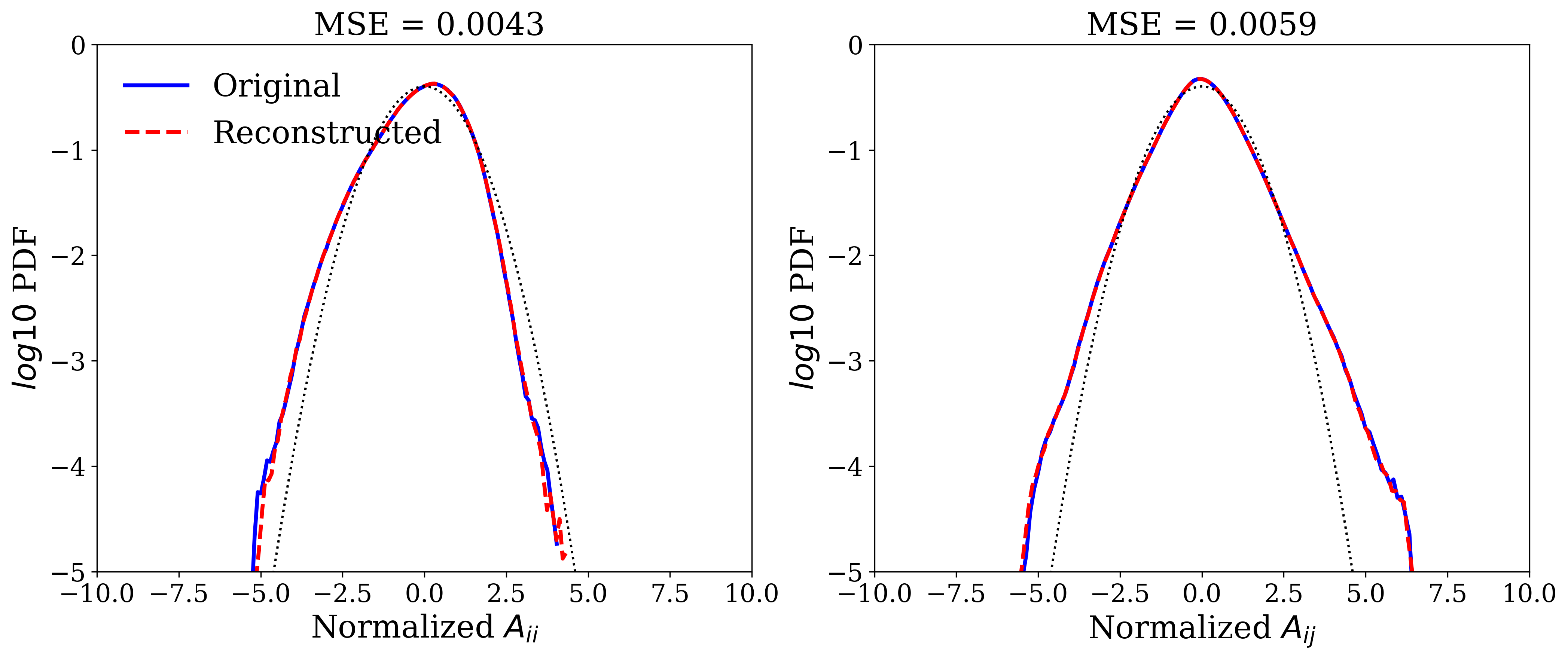}
		\caption{Large scales}
	\end{subfigure}
	\caption{Comparing original and reconstructed 3D DIT compressed by VQ-AE with $SF=2$ (compression ratio  $CR = 85$): PDFs of normalized longitudinal (diagonal) and transverse (off-diagonal) components of velocity gradient tensor $\boldsymbol{A}$. }\label{fig:DIT_VG_Turb_uvw_vqvae_1_exact}
\end{figure}

Results for the PDFs of the longitudinal and transverse components of the velocity gradient tensor in the DIT snapshot are shown in figure \ref{fig:DIT_VG_Turb_uvw_vqvae_1_exact}. Compared to the results in the study of \cite{glaws2020deep}, our model more accurately captures the tails of the PDFs, with minor deviations in the far tails which disappears when considering filtering lenghts in the inertial and integral scale ranges.

\FloatBarrier

\begin{figure}[h]
	\centering
	
		\includegraphics[width=\linewidth]{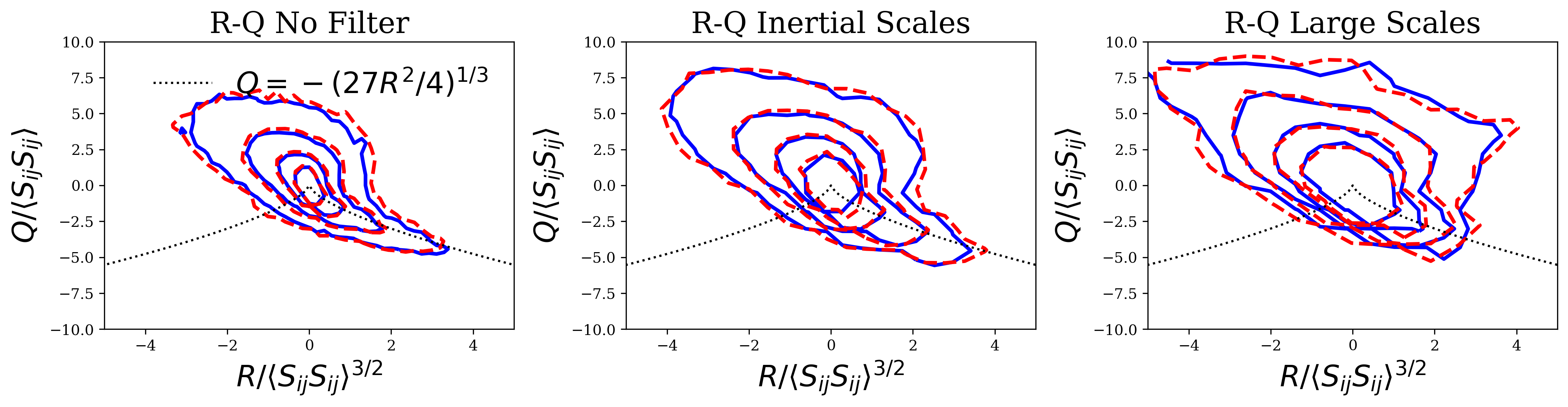}
	\caption{Comparing original and reconstructed 3D DIT compressed by VQ-AE with $SF=2$ (compression ratio  $CR = 85$): Contour plots of the joint PDF of normalized $Q=-Tr(\boldsymbol{A}^{2})/2$ and $R=-Tr(\boldsymbol{A}^{3})/3$. }\label{fig:DIT_RQ_Turb_uvw_vqvae_1_exact}
\end{figure}

The results for the joint-PDF of the $Q$ and $R$ invariants of velocity gradient tensor for the DIT flow are shown in figure \ref{fig:DIT_RQ_Turb_uvw_vqvae_1_exact}, which was not studied in \cite{glaws2020deep}. Just as for the HIT case, the tear-drop shape of the PDF is accurately captured with our model, despite providing a compression ratio of 85. 

\FloatBarrier

\subsection{Taylor-Green vortex}
In this section, we evaluate the performance of our model on a (fully-developed) realization from decaying Taylor-Green vortex (TGV) turbulence \cite{taylor1937mechanism}, which has been simulated following the procedure in \cite{brachet1984taylor}. The TGV flow is a popular turbulent flow for benchmarking CFD solvers as it has a (short-time) analytical solution. At longer times, three-dimensional vortex stretching leads to the generation of small-scales in the flow and turbulence, and the flow is non-isotropic (but still homogeneous), providing a good test of the model which was trained on an isotropic turbulent flow. Moreover, this TGV flow was simulated on a cubic domain with $192$ gird points in each direction, and so can also challenge our model with respect to its ability to be agnostic to the dimension of the input data.

In table \ref{tab:Summary_TGV_SF2_Table}, we summarize the results of the reconstructed flow field based on our model with $SF =2$ (corresponding to $CR=85$) which compresses the $3 \times 192^3$ floating point data into a discrete latent space and represent it with $1 \times 96^3$ integer data. Although we could not access the exact TGV flow snapshot used in the study of \cite{glaws2020deep}, to make a fair comparison we tested our model on multiple snapshots to verify that our model has a robust performance across different realizations. Compared to the results presented in \cite{glaws2020deep}, our results are almost the same as theirs with respect to the point-wise metrics (MSE and MAE), but we obtain around $10 \%$ lower $PSNR$ and $MSSIM$. In terms of physics-based metrics, the reconstructed terms for the quantities appearing in the first Betchov relationship (eq. \ref{eq:Betchov_1}) are reasonably accurate, although that relationship itself does not hold for the TGV flow. There are, however, large deviations for the higher-order terms in the second Betchov relation eq. \ref{eq:Betchov_2}, though the quality of the results improve considerably as we filter out the small scales of flow.

\begin{table}[]
\centering
\begin{tabular}{cccc}
\hline
Metrics                                                                                                                & ($SF=2$, Small scales) & ($SF=2$, Inertial scales) & ($SF=2$, Large scales) \\ \hline
MSE                                                                                                                    & $2.69\times10^{-3}$    & $2.13\times10^{-3}$       & $1.58\times10^{-3}$    \\ 
MAE                                                                                                                    & $3.95\times10^{-2}$    & $3.58\times10^{-2}$       & $3.15\times10^{-2}$    \\ 
PSNR                                                                                                                   & $26.5$                 & $25.82$                   & $25.52$                \\ 
MSSIM                                                                                                                  & $0.83$                 & $084$                     & $0.85$                 \\ 
$\langle S_{ij} S_{ij} \rangle$,$\langle \widehat{ S_{ij} S_{ij}} \rangle$                                             & $4.2\:,\:3.54$         & $2.5\:,\:2.1$             & $1.49\:,\:1.3$         \\ 
$\langle R_{ij} R_{ij} \rangle$,$\langle \widehat{ R_{ij} R_{ij}} \rangle$                                             & $2.43\:,\:2.35$        & $1.44\:,\:1.39$           & $0.859\:,\:0.824$      \\ 
$(-3/4)\times$ ($\langle S_{ij} \omega_{j} \omega_{j}\rangle$,$\langle \widehat{S_{ij} \omega_{j} \omega_{j}}\rangle$) & $-0.047\:,\:-0.101$    & $-0.024\:,\:-0.030$       & $-0.018\:,\:-0.017$    \\ 
$\langle  S_{ij} S_{kj} S_{ji} \rangle$,$\langle \widehat{ S_{ij} S_{kj} S_{ji}} \rangle$                              & $-0.685\:,\:0.1525$    & $-0.041\:,\:0.055$        & $-0.011\:,\:0.015$     \\ \hline
\end{tabular}	\caption{Summary of the performance of trained VQ-AE evaluated on an unseen data from decaying Taylor-Green vortex turbulence. $SF=2$ represents scaling the input data ($3 \times 192^{3}$) by a factor of two which yields a compressed integer representation with size $1 \times 96^{3}$ and $CR=85$.}{\label{tab:Summary_TGV_SF2_Table}}
\end{table}

\begin{figure}[h]
	\centering	
	\includegraphics[width=0.7\linewidth]{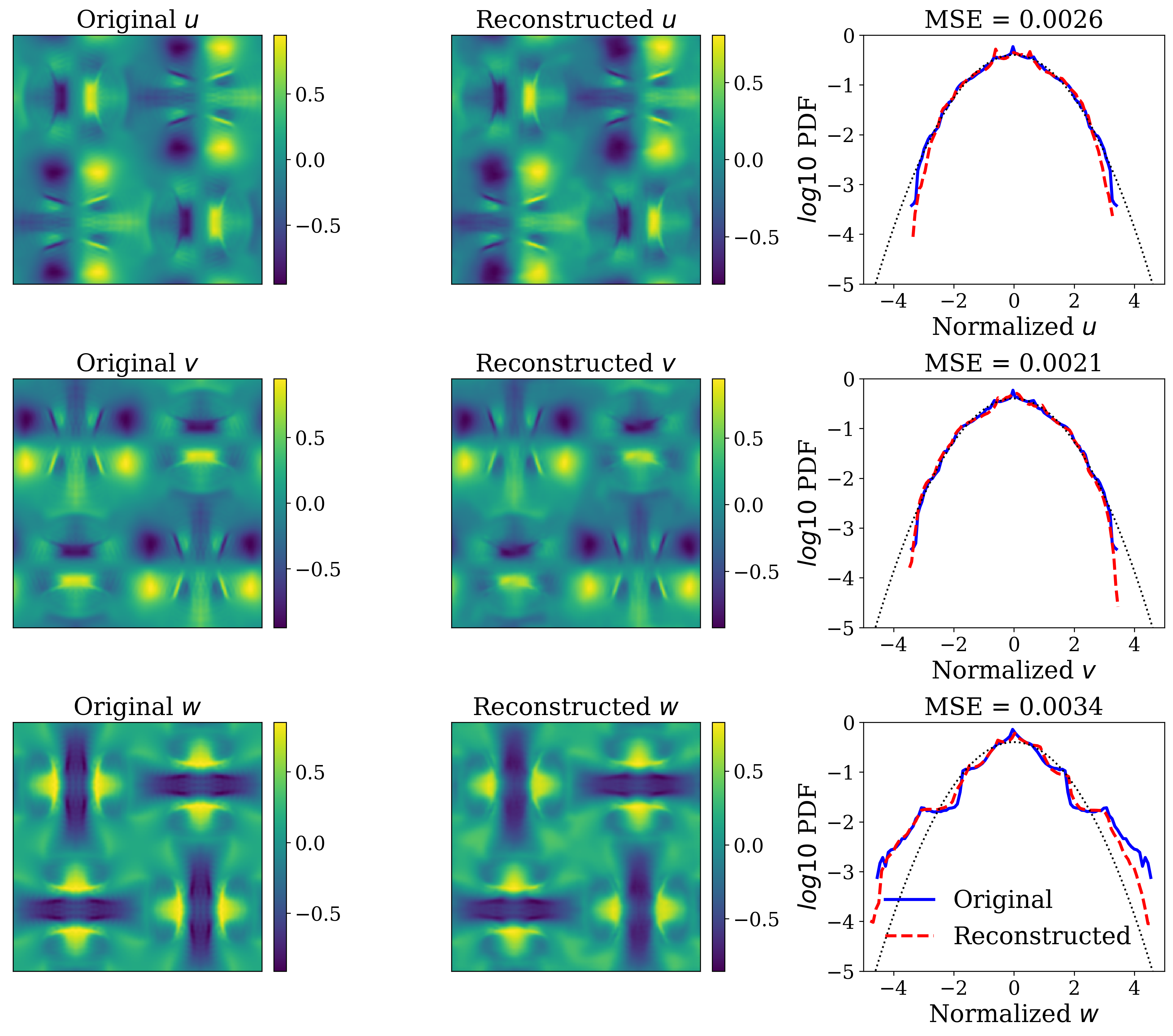}
	\caption{Comparing original and reconstructed 3D TGV compressed by VQ-AE with $SF=2$ (compression ratio  $CR = 85$): 2D snapshots and PDFs of velocity components}
	\label{fig:TGV_uvw_0_Turb_uvw_vqvae_1_exact}
\end{figure}

Visual comparison of two-dimensional snapshots of the velocity component fields from both the original and reconstructed fields, shown in figure \ref{fig:TGV_uvw_0_Turb_uvw_vqvae_1_exact}, illustrate that our VQ-AE model successfully captures all the characteristics of this TGV flow. This figure also shows that the PDFs of the reconstructed velocity field, computed based on all the grid points in the three-dimensional domain, match well with the original flow field, except for a minor deviation in the right tail of the $w$ PDF. In addition to the PDFs of the velocity components, the rest of the results in this section were not studied in \cite{glaws2020deep} and therefore we cannot compare the quality of our model with their with respect to the ability to capturing these statistical properties of the TGV flow.

\FloatBarrier

\begin{figure}[h]
	\centering	
	\includegraphics[width=0.4\linewidth]{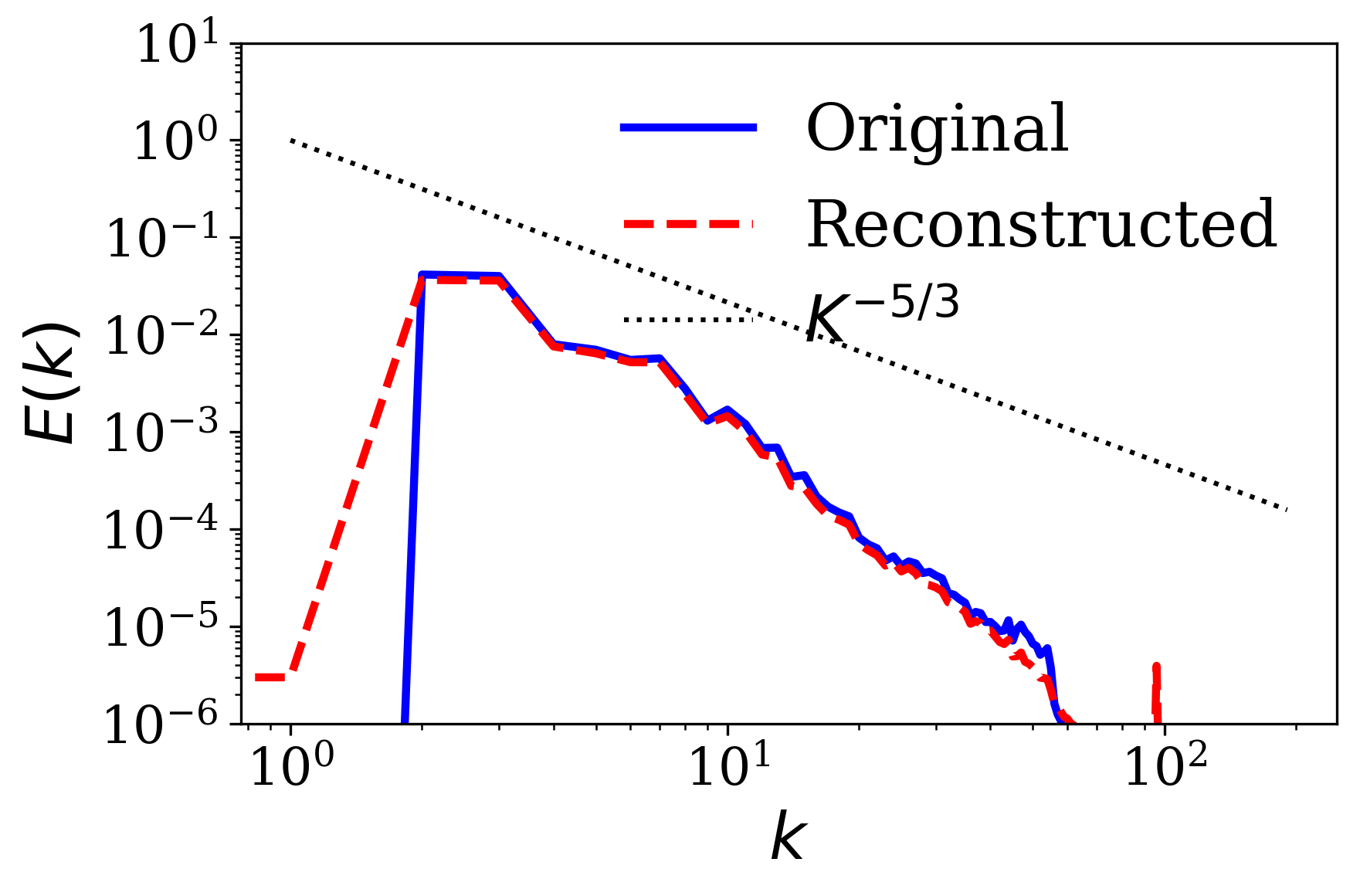}
	\caption{Comparing original and reconstructed 3D TGV compressed by VQ-AE with $SF=2$ (compression ratio  $CR = 85$): Turbulence Kinetic Energy Spectra }
	\label{fig:TGV_EnergySpectrum_0_Turb_uvw_vqvae_1_exact}
\end{figure}

Figure \ref{fig:TGV_EnergySpectrum_0_Turb_uvw_vqvae_1_exact} presents the comparison between TKE spectrum of the reconstructed and original TGV flow field. From this figure, it is apparent that our model can fully recover the power law behavior of energy spectrum across the inertial scales of flow and the deviations at the smallest scales are minimal. The deviation at the smallest wavenumbers (large scales) can be attributed to the fact that in the original data, the energy content at these scales is very small, $O(10^{-32})$, and so it is hard for the model to retain such precision in the reconstructed data.

\FloatBarrier

\begin{figure}[h]
	\centering
	\begin{subfigure}[b]{\linewidth}
		\includegraphics[width=0.85\linewidth]{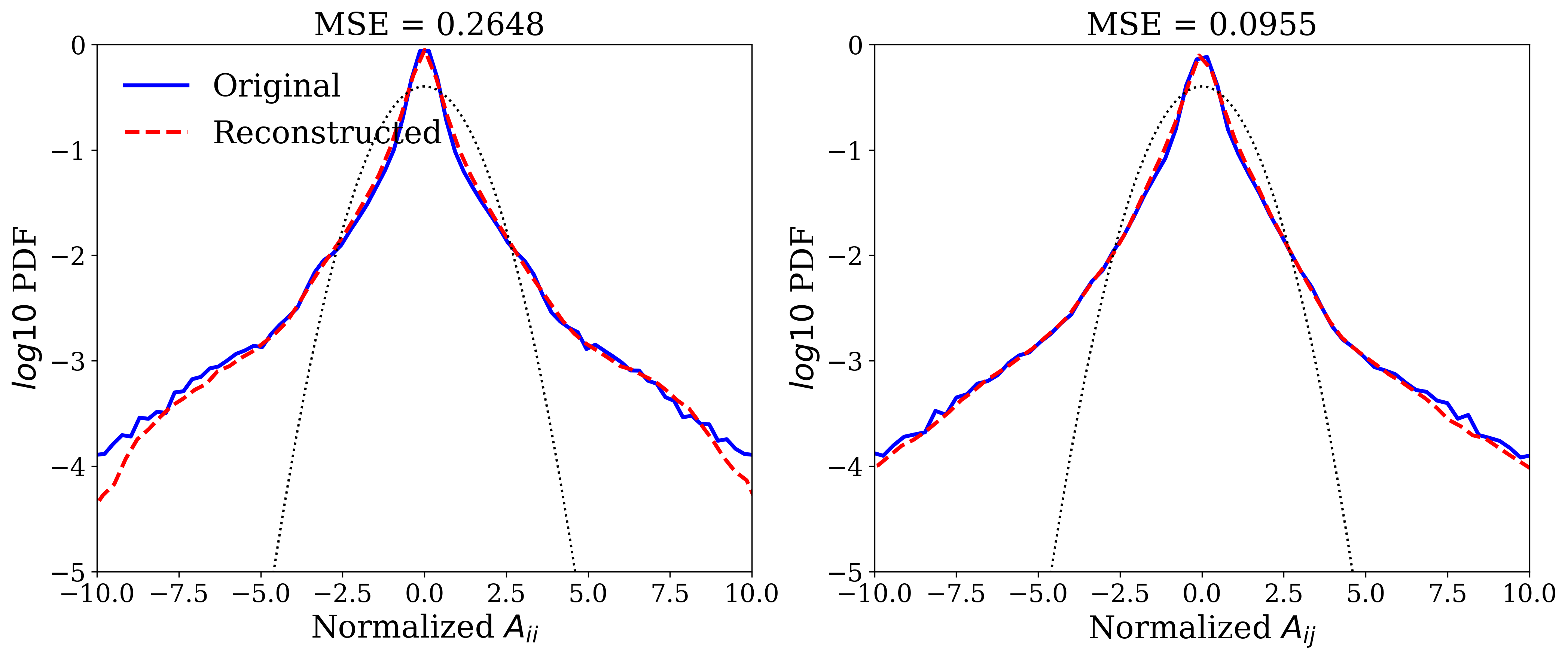}
		\caption{No filter}
	\end{subfigure}
	\begin{subfigure}[b]{\linewidth}
		\includegraphics[width=0.85\linewidth]{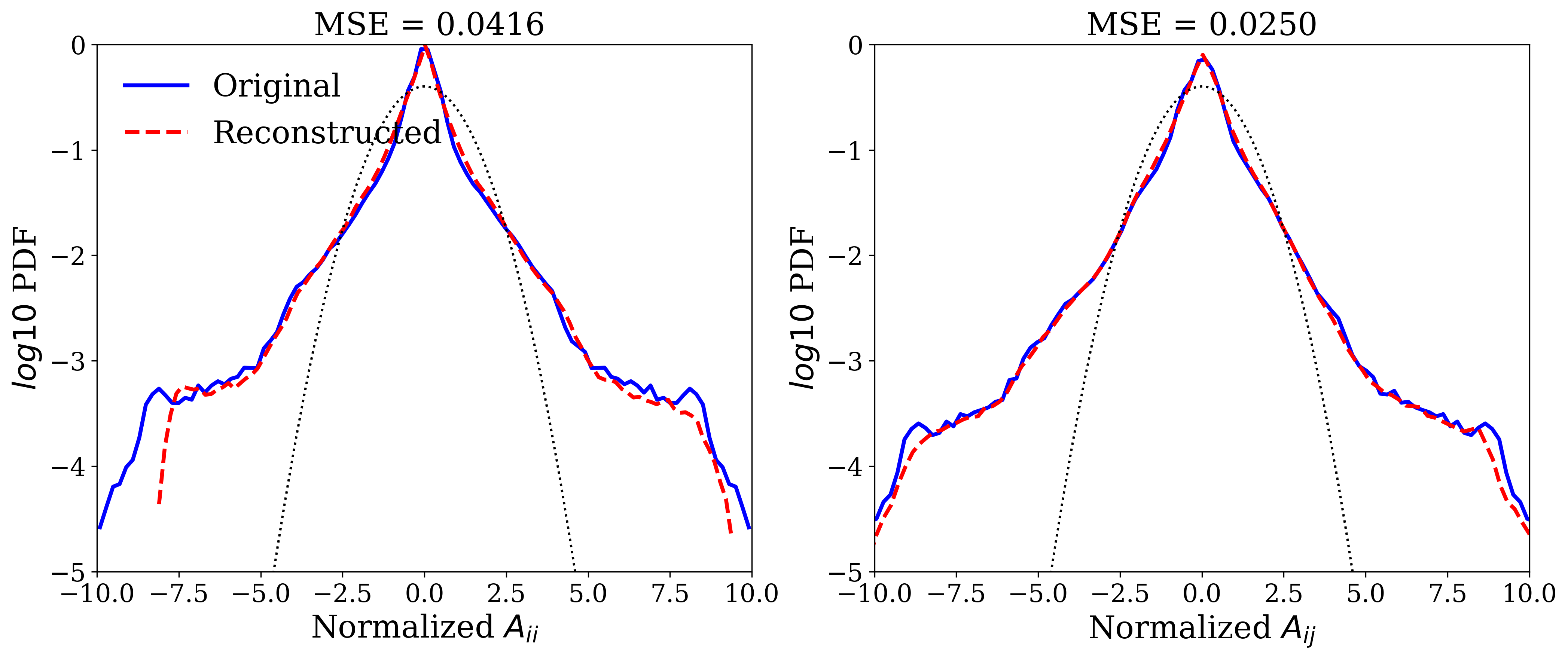}
		\caption{Inertial scales}
	\end{subfigure}
		\begin{subfigure}[b]{\linewidth}
		\includegraphics[width=0.85\linewidth]{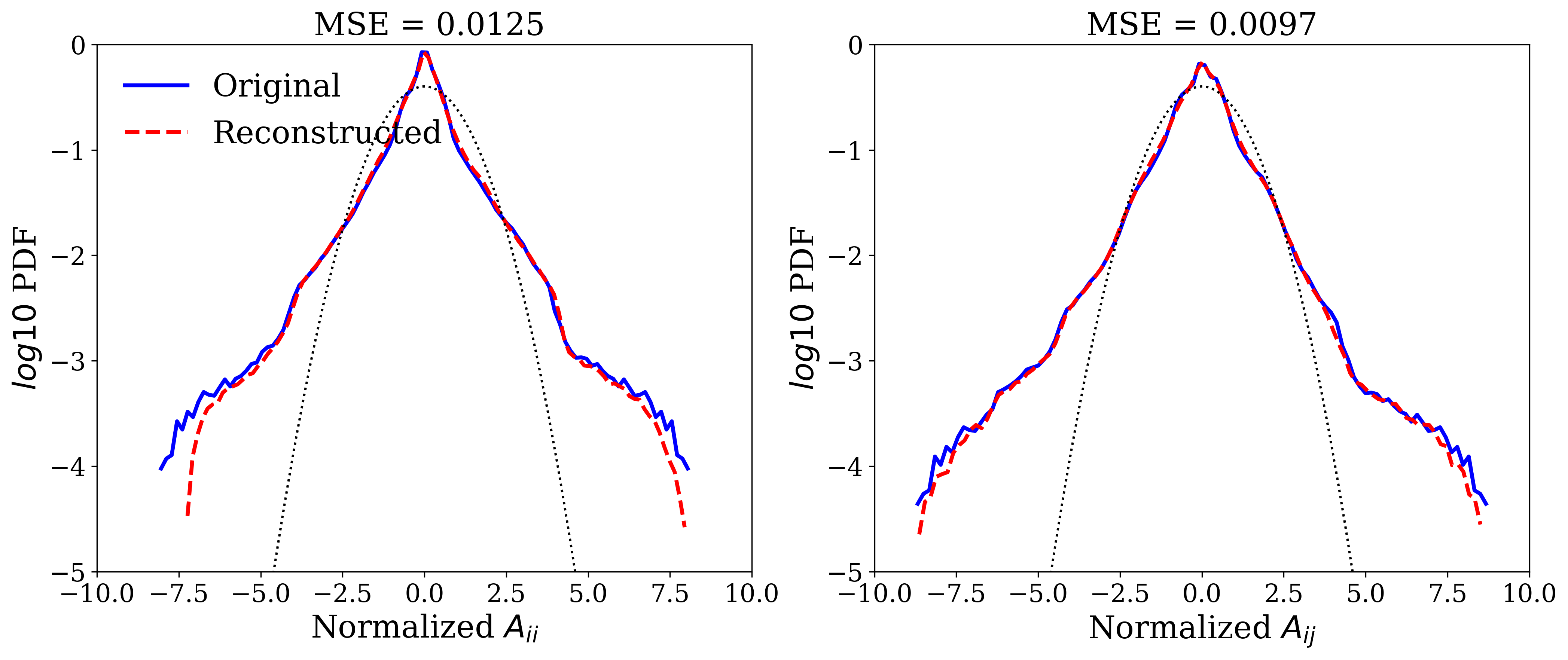}
		\caption{Large scales}
	\end{subfigure}
	\caption{Comparing original and reconstructed 3D TGV compressed by VQ-AE with $SF=2$ (compression ratio  $CR = 85$): PDFs of normalized longitudinal (diagonal) and transverse (off-diagonal) components of velocity gradient tensor $\boldsymbol{A}$. }\label{fig:TGV_VG_Turb_uvw_vqvae_1_exact}
\end{figure}

Figure \ref{fig:TGV_VG_Turb_uvw_vqvae_1_exact} shows the PDFs of the longitudinal and transverse components of the velocity gradient tensor in the TGV flow. It can be seen that our model performs extremely well, both qualitatively and quantitatively, accurately capturing both the body and tails of the PDFs of this anisotropic flow, with minor errors in the far tails of longitudinal components. Just as already observed for the previous test data, this error mainly stems from smallest scales of flow and disappears when filtering out these scales.

\FloatBarrier

\begin{figure}[h]
	\centering
	
		\includegraphics[width=\linewidth]{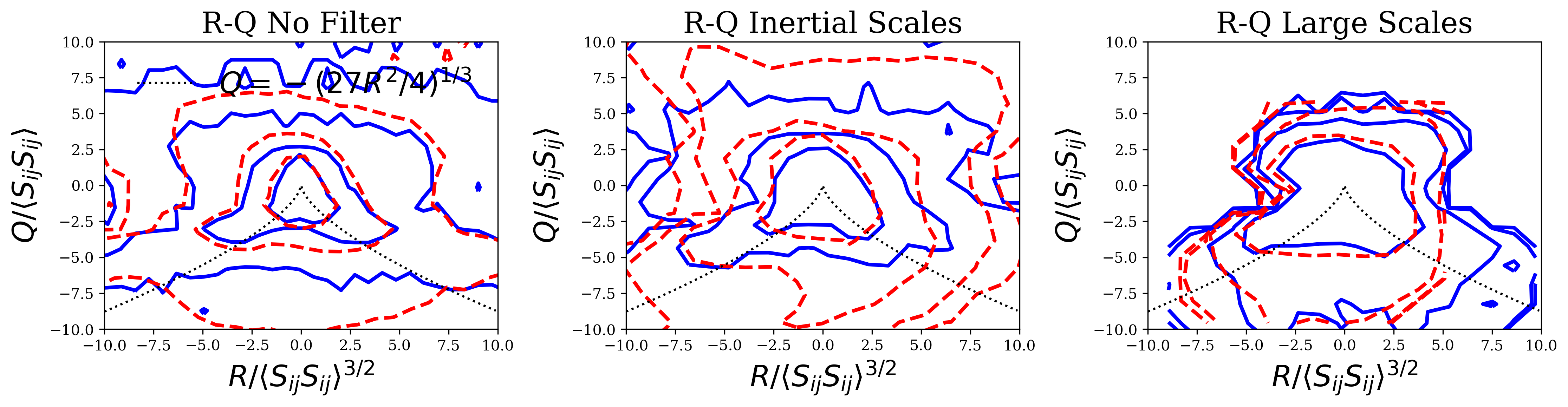}
	\caption{Comparing original and reconstructed 3D TGV compressed by VQ-AE with $SF=2$ (compression ratio  $CR = 85$): Contour plots of the joint PDF of normalized $Q=-Tr(\boldsymbol{A}^{2})/2$ and $R=-Tr(\boldsymbol{A}^{3})/3$. }\label{fig:TGV_RQ_Turb_uvw_vqvae_1_exact}
\end{figure}

In figure \ref{fig:TGV_RQ_Turb_uvw_vqvae_1_exact} we show the results for the joint-PDF of the $Q$ and $R$ invariants of velocity gradient tensor for this TGV flow. Unlike the previous isotropic flow test cases, our model can only capture the most frequent characteristics of the flow (interior contours of the PDF), and fails to recover the less frequent ones. This may seem surprising given that the model performed so well in predicting the PDFs of the components of the velocity gradients. However, the invariants $Q,R$ depend not only upon the properties of the individual velocity gradient components, but also upon more subtle features such as the geometric alignments between the strain-rate and vorticity fields \cite{meneveau11,tsinober}. Another point is that even for the original data, the PDFs are very noisy compared to the earlier isotropic test cases. This significant statistical noise may be another reason why the model struggles to fully capture the behavior of the $Q,R$ PDF for the TGV flow.      

\FloatBarrier

\subsection{Channel flow}

In Table \ref{tab:Summary_ChannelFlow_Table} we summarize our results for the performance of the trained VQ-AE on a flow snapshot from a fully developed, turbulent channel flow. This is a completely different flow than that which our model was trained on, being strongly inhomogeneous and anisotropic, and therefore constitutes a good test for the general applicability of our model. While in the previous sections we presented only the results of our model with the regularization terms ($\alpha=0.1$, $\gamma=10^{-4}$), here we also report the results for the case without the regularization terms ($\alpha=\gamma=0$). As we will show, overall we found that for the channel flow, our VQ-AE model performs slightly better without regularization terms than with. This is most likely related to the fact that those constraints do not apply to this type of flow, but only to statistically homogeneous flows. As noted earlier when considering the HIT results, however, incorporation of the regulariztion terms only slightly improved the model predictions for that case. Therefore, the version of our model without the regularization terms seems to provide a model that is optimally accurate across a range of different flow types.

Compared to the results presented in \cite{glaws2020deep}, whose fully convolutional AE model provides $CR=64$, our $SF=2$ model (with regularization) improves both the $MSE$ and $MAE$ by an order of magnitude, but yields a 15\% lower $PSNR$, and a 10\% lower $MSSIM$ than their model. However, our model for $SF=2$ improves on their compression ratio by more than 30\%.

\begin{table}[]
\centering
\begin{tabular}{ccccc}
\hline
                                 & MSE                & MAE                 & PSNR   & MSSIM  \\ \hline
$SF = 2$, with regularization    & $3.0\times10^{-2}$ & $1.29\times10^{-1}$ & $24.8$ & $0.83$ \\
$SF = 2$, without regularization & $2.4\times10^{-2}$ & $1.09\times10^{-1}$ & $26.2$ & $0.82$ \\ \hline
\end{tabular}
\caption{Summary of the performance of trained VQ-AE evaluated on a data from channel flow. $SF=2$ represents scaling the input data ($3 \times 128^{3}$) by a factor of two which yields a compressed integer representation with size $1 \times 64^{3}$ and $CR=85$. Model with no regularization means $\alpha=\gamma=0$.}
	{\label{tab:Summary_ChannelFlow_Table}}
\end{table}

\begin{figure}[h]
	\centering
	\begin{subfigure}[b]{\linewidth}
		\includegraphics[width=0.6\linewidth]{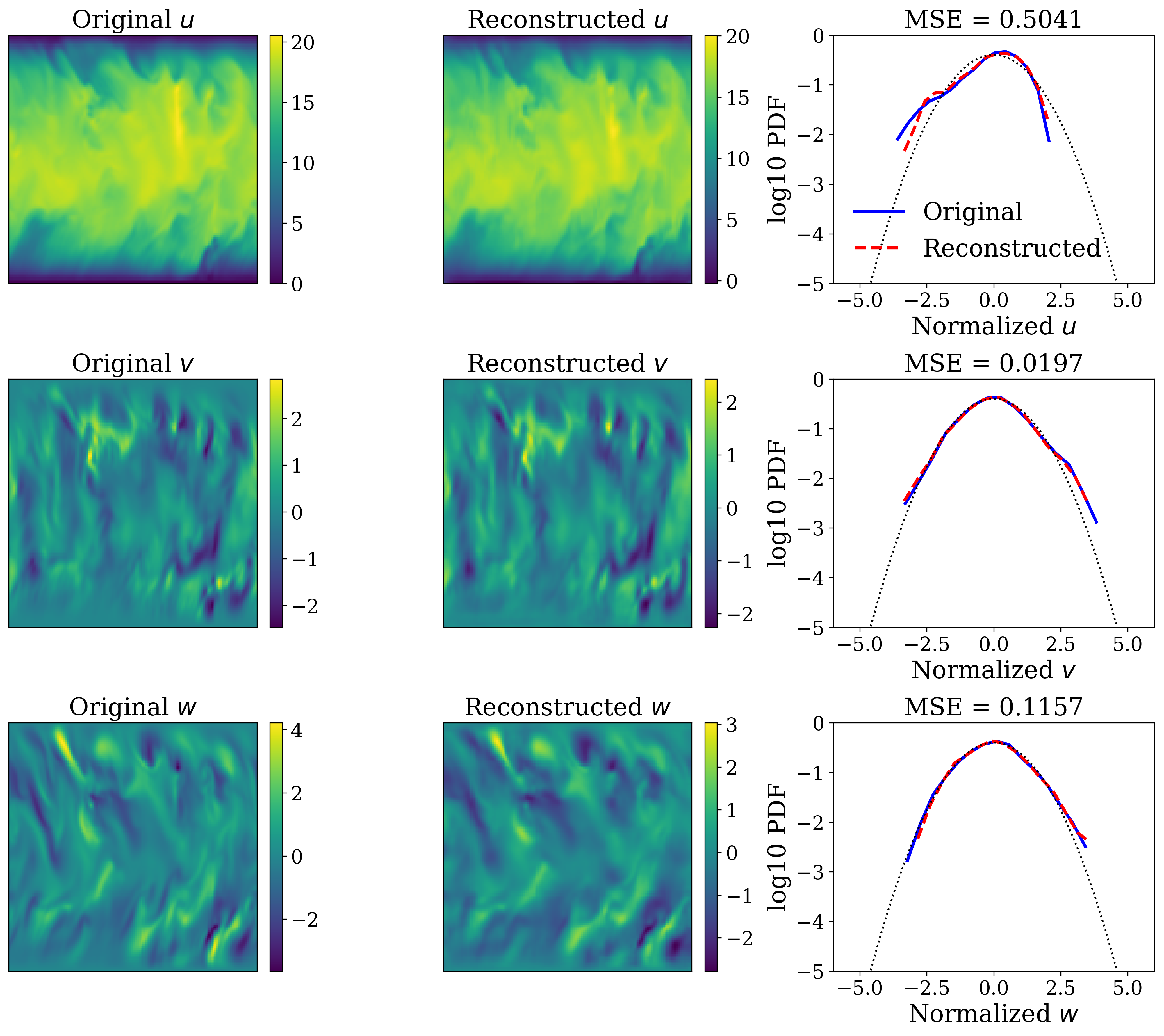}
		\caption{With Regularization}
	\end{subfigure}
	\begin{subfigure}[b]{\linewidth}
		\includegraphics[width=0.6\linewidth]{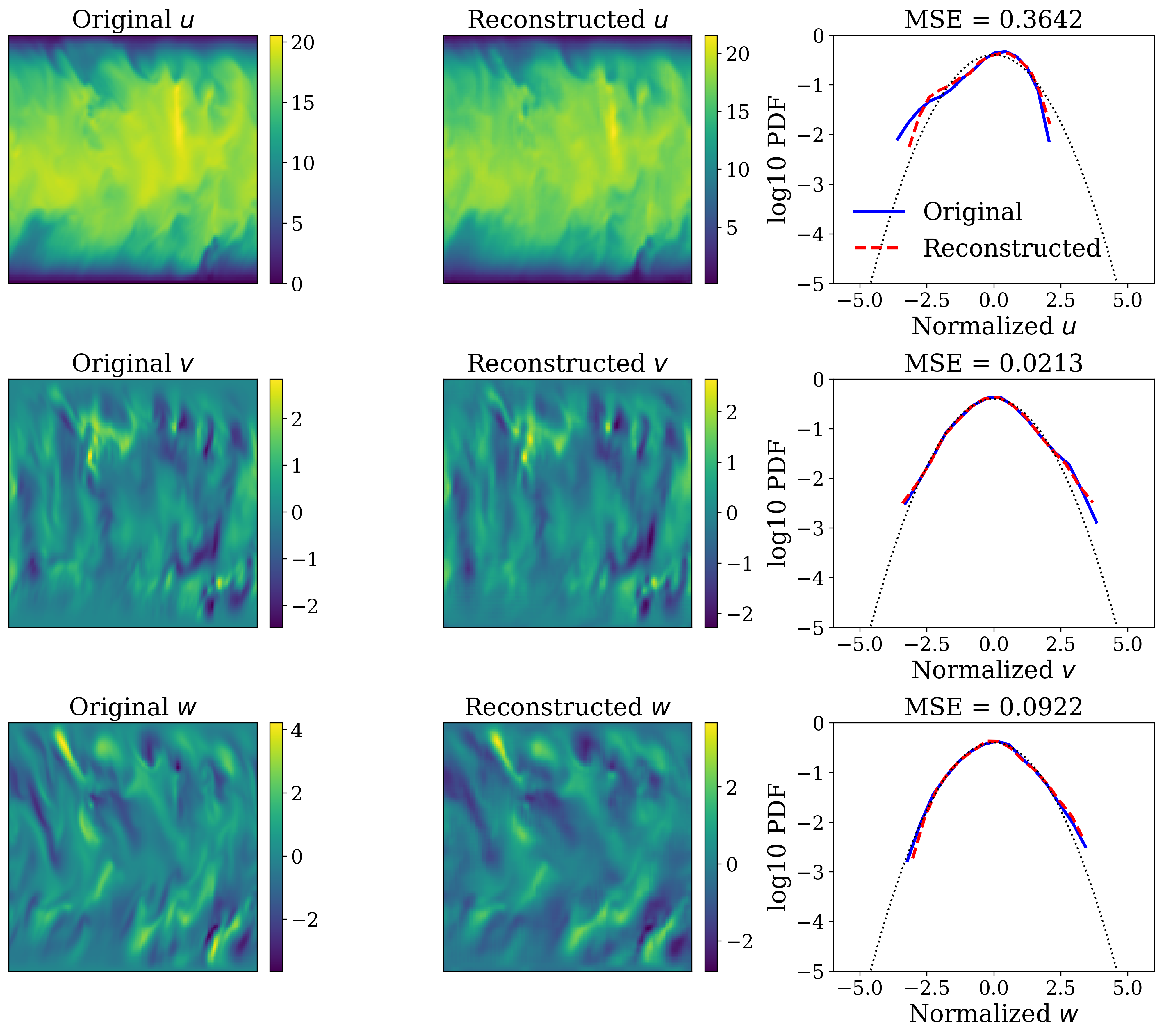}
		\caption{Without Regularization}
	\end{subfigure}
	 \caption{Comparing original and reconstructed 3D channel flow compressed by vqvae with $SF=2$ (compression ratio  $CR = 85$): 2D snapshots and PDFs of velocity components}
	\label{fig:ChannelFlow_uvw_0_Turb_uvw_vqvae_1_exact} 
\end{figure}

A visual comparison of the original and reconstructed channel flow snapshots is given in figure \ref{fig:ChannelFlow_uvw_0_Turb_uvw_vqvae_1_exact}, illustrating that our model performs well in capturing particular characteristics of the channel flow, including the no slip condition near walls, which were not present in the training data for the model. In this figure we also plotted the PDFs of velocity components at a fixed wall distance which illustrate that our model can fully capture distribution of the spanwise ($w$) and wall-normal ($v$) components but has some deviation at the left tail of the streamwise ($u$) component.

\FloatBarrier

\begin{figure}[h]
	\centering

		\includegraphics[width=1\linewidth]{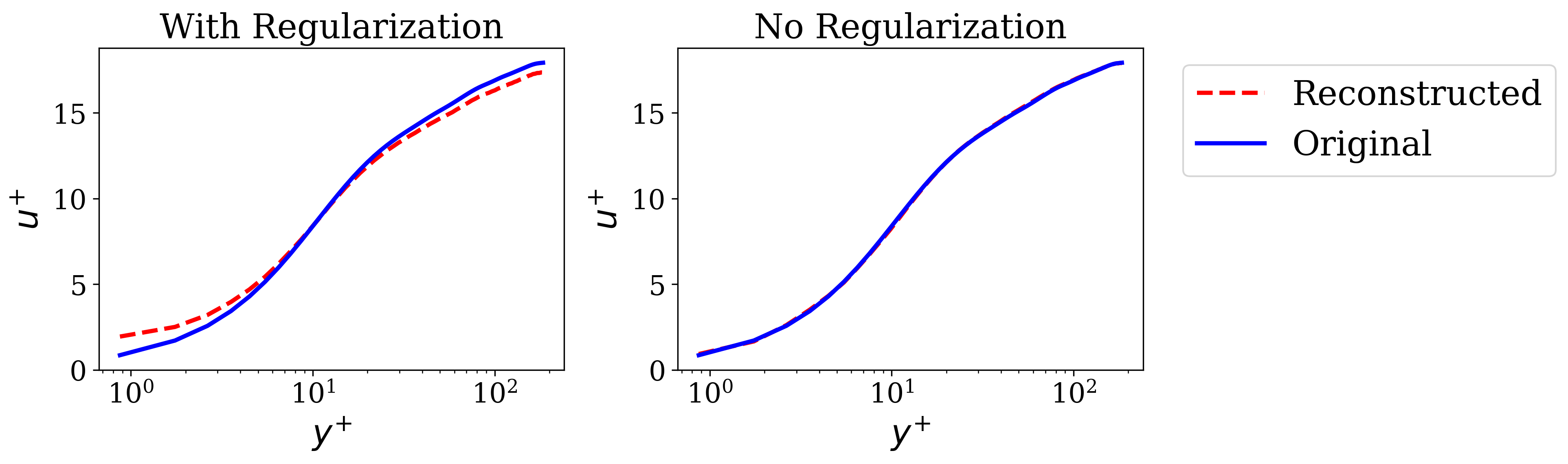}
	
	\caption{Comparing original and reconstructed 3D channel flow compressed by VQ-AE with $SF=2$ (compression ratio  $CR = 85$): Averaged (along streamwise and spanwise directions) streamwise velocity as a function of the normalized wall distance.}\label{fig:u+_ChannelFlow_Turb_uvw_vqvae_1_exact}
\end{figure}

In figure \ref{fig:u+_ChannelFlow_Turb_uvw_vqvae_1_exact} we plot $u^{+}=\langle u \rangle/u_{\tau}$, the average streamwise velocity normalized by the friction velocity $u_\tau$, as a function of the normalized wall distance, $y^{+}=u_{\tau} y/\nu$. We see that including the regularization terms leads to significant errors in the model predictions, which as mentioned before, is likely due to the fact that the regularization terms are based on results for homogeneous turbulent flows, and do not apply in channel flows. However, the model without the regularization terms reproduces the original DNS data almost perfectly.

\FloatBarrier

\begin{figure}[h]
	\centering

		\includegraphics[width=1\linewidth]{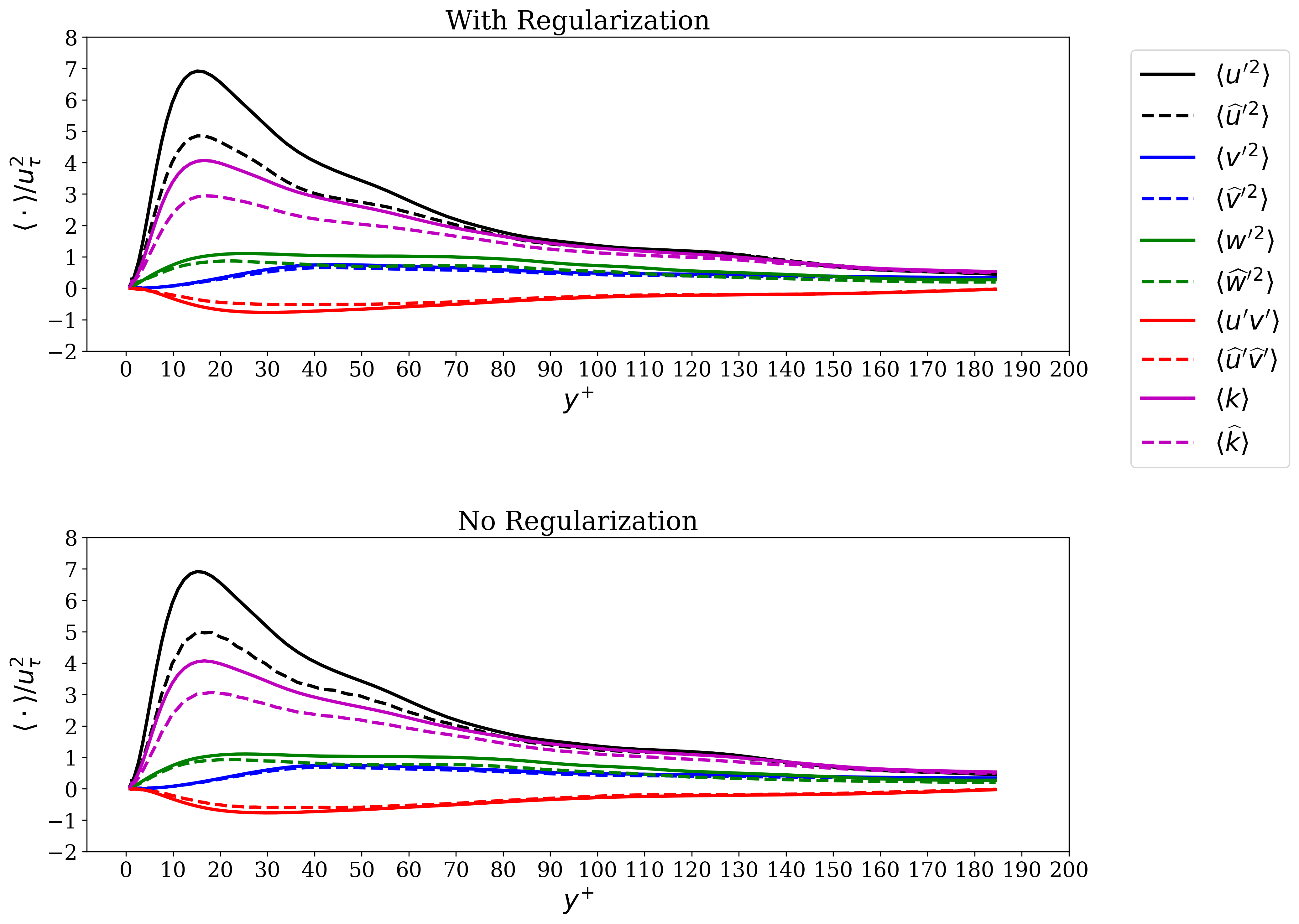}
	\caption{Comparing original and reconstructed 3D channel flow compressed by VQ-AE with $SF=2$ (compression ratio  $CR = 85$): Averaged (along streamwise and spanwise directions) Turbulent kinetic energy and Reynolds stresses as a function of the normalized wall distance.}\label{fig:Other_ChannelFlow_Turb_uvw_vqvae_1_exact}
\end{figure}

In figure \ref{fig:Other_ChannelFlow_Turb_uvw_vqvae_1_exact} we plot the results for the flow Reynolds stresses and TKE from the original DNS and the model predictions. In the plot, $u', v', w'$ denote the fluctuating velocity components (i.e. the velocity with the mean value subtracted) in the streamwise, wall-normal, and spanwise directions, and the TKE is defined as $k \equiv ({u'}^2+{v'}^2+{w'}^2)/2$. Compared to the same results presented in \cite{glaws2020deep}, our model more accurately reconstructs all of the quantities in figure \ref{fig:Other_ChannelFlow_Turb_uvw_vqvae_1_exact} except the streamwise Reynolds stress, $\langle u'^{2}\rangle$. Again we observe that the model without regularization terms yields a better performance. More specifically, these results show that
our model fails to capture $\langle {u}'^2 \rangle$ in the range $5 < y^+ < 140$, with a maximum error of around $25\%$ at $y^+=15$. For the Reynolds stresses in the spanwise and wall-normal directions, our model has some error for $\langle {w}'^2 \rangle$ in the range $19 < y^+ < 130$, but fully captures $\langle {v}'^2 \rangle$ across the entire wall-normal direction. These deviations in the $\langle {u}'^2 \rangle$ and $\langle \widehat{w}'^2 \rangle$ result in errors in the TKE $\langle {k} \rangle$ in the range $5 < y^+ < 90$, with the largest error occurring at $y^+=15$. For $\langle {u}'{v}' \rangle$, the deviation is very small and occurs in the range $9 < y^+ < 50$. Overall, these results indicate the greatest errors in the model predictions occur in the buffer layer $5 < y^+ < 30$. This is not surprising since this is a region where flow anisotropy and inhomogeneity are strong, properties not present in the training data. It would be interesting in future work to see if including snapshots of a canonical flow such as a turbulent channel flow during the training process of the model would enable the model to accurately capture the properties of other strongly inhomogeneous and anisotropic turbulent flows.

\FloatBarrier
\vskip -0.5 cm
\section{Conclusions}\label{Conclusions}
\vskip -0.5 cm
With growing the scale of CFD simulations, the computational costs of handling the large simulation data files they produce becomes prohibitive in terms of storage and memory band-width. This highlights the importance of employing data compression as a tool to facilitate checkpointing, transferring, and post-processing such large scale data.

In this study, we have proposed a vector quantized deep learning framework, the so called vector quantized autoencoder or VQ-AE, for the compression of data from turbulent flow simulations. We have calibrated the loss function of the model to infuse prior knowledge of the flow in the form of constraints in order to boost the model performance.  

This lossy compression model encodes the velocity field data in a discrete latent space, and offers a minimum compression ratio of $85$. The resulting compressed data still accurately captures many of the key flow properties of the original data across all the scales of flow, and any deviations from the original data are mainly confined to the properties of the smallest scales of the flow. Compared to the recent data compression study of \cite{glaws2020deep}, in which they proposed a convolutional autoencoder which compresses the data in a continuous latent space with compression ratio of $64$, our model not only improves the compression capability by more than $30 \%$ but also reconstructs the small scales of the flow with much higher accuracy than theirs. Furthermore, our model is much cheaper from a computational perspective, with less than $1.4$ million parameters, and is trained with only 40 realizations of flow snapshot data, and can be easily trained on a single (NVIDIA Pascal P100) GPU in around 8 hours (while the maximum memory consumption is around 5 $ GB$).     

Our fully convolutional deep learning model has been trained on high fidelity DNS data of statistically stationary homogeneous isotropic turbulence (HIT), and its performance, based on a range of conventional metrics for image processing tasks and the physics of turbulence, has been evaluated on snapshots (one from each) from four different turbulent flows (i) unseen data from the same HIT simulation that the model was trained on, (ii) decaying isotropic turbulence (DIT), (iii) decaying Taylor-Green vortex (TGV), and (iv) pressure-driven channel flow. With respect to the HIT test case, which was not a challenging task for the model as the data has similar characteristics as the training set, our model fully recovered all the flow characteristics up to second order statistics of the velocity gradient tensor, with small discrepancies at the smallest scales. We also found that the embedding physics constraints in the loss function can noticeably improve the quality of the reconstructed small scales of the flow. For the DIT test case, in which we used the same data as in \cite{glaws2020deep}, we obtained the same level of accuracy as the HIT case and compared to \cite{glaws2020deep} our model improves $MSE$ by an order of magnitude, and can recover many of the small scale properties quantified by the turbulent kinetic spectrum and PDFs of the velocity gradient tensor. Regarding the TGV test case, which comes from a simulation on a $192 ^3$ domain (as opposed to the training set which simulated on a $128^3$ grid), and has considerably different characteristics to the flow used for training the model, our model recovers the velocity field with around $10 \%$ lower $PSNR$ and $MSSIM$ compared to the study of \cite{glaws2020deep} but captures the PDFs of the velocity components with very small distortion. Our model captures the PDFs of the components of the velocity gradient tensor with great accuracy (except some small deviation at the tails of PDFs). However, our model does not perform so well for capturing the propeties of the full velocity gradient invarient PDFs which are more subtle, as they are influenced by geometrical alignments between the strain-rate and vorticity fields in the flow. In the HIT, DIT and TGV cases, we found that most of the information loss due to compression is associated with the small scales, such that when those scales are filtered out using a low-pass Gaussian filter, the discrepancy between the original and reconstructed data disappears.

Finally, we tested our model on a realization from a turbulent channel flow, which is not only anisotropic but also inhomogeneous, and therefore very different from the HIT training data. We found that the model has satisfying performance in recovering important statistics of the flow, and the quality of results are comparable (even slightly better for some quantities) compared with the model of \cite{glaws2020deep}, except for the prediction of the streamwise Reynolds stress. Interestingly we found that our model without regularization terms included performs better for this channel flow case and we attribute this to the fact that the regularization constraints in the loss function are based on propeties for homogeneous turbulent flows, and therefore do no apply in a channel flow near the wall.  

Our data compression model can facilitate the dynamical modeling of three dimensional turbulence as it provides a latent space with high information content that can be used in a sequence modeling deep learning framework to learn the spatio-temporal characteristics of flow. Furthermore, we believe this data compression framework is not limited to CFD simulations but can be easily applied to compress data from other complex physical simulations with structured mesh domains.    

\vskip -0.5 cm
\section{Acknowledgments}
\vskip -0.5 cm
We would like to thank Michele Iovieno and Maurizio Carbone for kindly providing some of the DNS data used in this paper. This work used the Extreme Science and Engineering Discovery Environment (XSEDE), which is supported by National Science Foundation grant number ACI-1548562 \cite{xsede}. Specifically, the Comet cluster was used under allocation CTS170009 and the authors would like to thank Marty Kandes for his assistance with setting up the GPU environment. This work was also supported by the Office of Naval Research (ONR) under grant number N00014-18-1-2244.

\bibliographystyle{unsrt}
\bibliography{main}


\end{document}